\documentclass{blois}
\pdfoutput=1 
\usepackage{jinstpub} 
\notoc

\usepackage[colorinlistoftodos]{todonotes}
\usepackage{HGTD_TDR-defs}
\usepackage{multirow}
\usepackage{rotating}
\usepackage{mathrsfs}
\usepackage{booktabs}
\usepackage{longtable}
\usepackage{placeins}
\usepackage{amssymb}
\usepackage{upgreek}
\usepackage{svg}
\usepackage{comment}
\usepackage{pdfpages}
\usepackage{siunitx}
\usepackage{lineno}
\usepackage{float}
\usepackage{lineno}
\usepackage{comment}
\usepackage{flafter}
\usepackage{subcaption}

\newcommand{\tmax}{{t_\text{max}}}
\newcommand{\pmax}{{p_\text{max}}}

\title{\boldmath Synchrotron light source X-ray detection with Low-Gain Avalanche Diodes}

\abstract{
Low Gain Avalanche Diodes (LGADs) represent the state-of-the-art in timing measurements and will instrument the future Timing Detectors of ATLAS and CMS for the High-Luminosity LHC. While initially conceived as a sensor for charged particles, the intrinsic gain of LGADs makes it possible to detect low-energy X-rays with good energy resolution and excellent time resolution (tens of picoseconds). Using the Stanford Synchrotron Radiation Lightsource (SSRL) at SLAC, several LGADs designs were characterized with energies from 5 to 70 \si{\kilo\electronvolt}. The SSRL provides \SI{10}{\pico\second} pulsed X-ray bunches separated by \SI{2}{\nano\second} intervals with an energy dispersion ($\Delta E/E$) of $10^{-4}$. LGADs from Hamamatsu Photonics (HPK) and Brookhaven National Laboratory (BNL) with different thicknesses ranging from \SI{20}{\micro\meter} to \SI{50}{\micro\meter} and different gain layer designs were read out using fast amplification boards and digitized with a high bandwidth and high sampling rate oscilloscope. PIN devices from HPK and AC-LGADs from BNL were characterized as well. A systematic and detailed characterization of the devices’ energy linearity, resolution, and time resolution as a function of X-ray energy was performed for different biasing voltages at room temperature and are reported in this work. The charge collection and multiplication mechanism were simulated using Geant4 and TCAD Sentaurus, providing an important handle for interpreting the data.}

\author[1]{S.M. Mazza}
\author[2]{G.~Saito}
\author[1]{Y.~Zhao}
\author[1]{T.~Kirkes}
\author[1]{N.~Yoho}
\author[1]{D.~Yerdea}
\author[1]{N.~Nagel}
\author[1]{J.~Ott}
\author[1]{M.~Nizam}
\author[2]{M.~Leite}
\author[4]{M.~Moralles}
\author[1]{H.~F.-W.~Sadrozinski}
\author[1]{A.~Seiden}
\author[1]{B.~Schumm}
\author[1]{F.~McKinney-Martinez}
\author[3]{G.~Giacomini}
\author[3]{W.~Chen}

\affiliation[1]{SCIPP, University of California Santa Cruz, 1156 high street, Santa Cruz (CA), US}
\affiliation[2]{Universidade de São Paulo, São Paulo (SP), Brazil}
\affiliation[3]{Brookhaven National Laboratory, Upton (NY), U.S.}
\affiliation[4]{IPEN-CNEN,  São Paulo (SP), Brazil}

\emailAdd{simazza@ucsc.edu}


\arxivnumber{2306.15798} 

\begin{document}







\maketitle
\flushbottom

\section{Introduction}
\label{sec:intro}

Several groups developing Low Gain Avalanche Detectors (LGADs) for use in  particle physics applications (e.g., timing layers for ATLAS and CMS~\cite{CERN-LHCC-2020-007,CMS:2667167}) have recently begun to explore the possibility of their application to other fields of science and technology~\cite{GALLOWAY20195}.
LGADs are silicon sensors with moderate internal gain and tens of picoseconds time resolution~\cite{bib:LGAD,bib:UFSD300umTB}. They comprise a low-doped region called 'bulk', typically 20-50~\si{\micro\meter} thick for fast timing applications (e.g. 50~\si{\micro\meter} for the ATLAS/CMS timing layers), and a highly doped thin region, a few \si{\micro\meter} from the charge collection electrode, called 'gain layer'. Thin LGADs have a fast rise time, exceptional time resolution, and short full charge collection time.

The response of LGADs to X-rays of energies 5-70 \si{\kilo\electronvolt} (with a  $\Delta E/E$  of $10^{-4}$) was characterized at the Stanford Light Source (SSRL)~\cite{SSRL} at the beamline 11-2~\cite{SSRL112}. SSRL 11-2 has a nominal beam size of $\SI{25}{\milli\meter}\times\SI{1}{\milli\meter}$ and a repetition rate about \SI{500}{\mega\hertz}. The tested thin LGADs easily resolved in time the repetition rate of the beam line.
The energy and time resolution of the tested LGADs were measured as a function of X-ray energy, device type, and gain. 
The characterization presented in this paper will be increasingly important since emerging prototypes (such as AC-LGADs~\cite{Heller:2022aug}, TI-LGADs~\cite{9081916}, DJ-LGADs~\cite{Ayyoub:2021dgk}) can reduce the LGAD spatial resolution to the scale of \SI{50}{\micro\meter} or less, allowing the technology to be reliably used for X-ray detection.
The tested devices are introduced in Sec.~\ref{sec:devices} and the experimental setup in the beam line is described in Sec.~\ref{sec:setup}. 

Simulation software was employed to understand the behavior of X-ray interaction with the sensors. Geant4 was used to evaluate the interaction rate  in the devices. TCAD Sentaurus was used to understand the charge collection mechanism for X-ray interaction in different areas of the LGAD tested. The Geant4 and TCAD simulations are described in Sec.~\ref{sec:geant4} and Sec.~\ref{sec:TCAD}, respectively.
An in-depth visualization of the average pulses is presented in Sec.~\ref{sec:pulses}. The data analysis procedure and results for the three standard LGADs and PIN are shown in Sec.~\ref{sec:analysis}, the results on the AC-LGAD sensors are presented in Sec.~\ref{sec:ACLGAD}.

\section{X-ray detection with LGADs}
\label{sec:xray_lgads}
This paper fits in well with an ongoing effort from several groups to introduce the LGAD technology to the photon physics community~\cite{iworid1,iworid2}.
In the field of X-ray detection for the diffraction field and medical imaging, the rates will be tied to those of the next-generation machines (the upgraded Argonne APS and then proposed next-generation XFELs, for example). 
For these, granularity requirements are on the order of 50x50~\SI{}{\micro\meter}, and the repetition rate will be over the GHz.
In this type of application, the internal gain of the LGADs multiplies the charge deposited by the X-rays to boost the signal-to-noise ratio, allowing the detection of low-energy X-rays, while simultaneously enabling the rapid signal collection for an ultra-high frame-rate response. 
This will enable X-ray cameras with frame rates above 100MHz as a possible application of high-granularity LGADs. 
In the longer term, LGADs can be envisioned to contribute to next-generation X-ray imaging and diffraction studies, which are expected in turn to be central to the development of products and techniques in a broad range of applied fields, including new materials, drug design, advanced computing, and homeland security, among others. 
Therefore, the characterization of standard LGAD devices and high-granularity LGAD devices (e.g., AC-LGADs) with soft X-rays is of great interest to the scientific community.

\section{Devices tested}
\label{sec:devices}
\begin{figure}[H]
\begin{subfigure}{0.25\textwidth} 
\centering
\includegraphics[width=\textwidth]
        {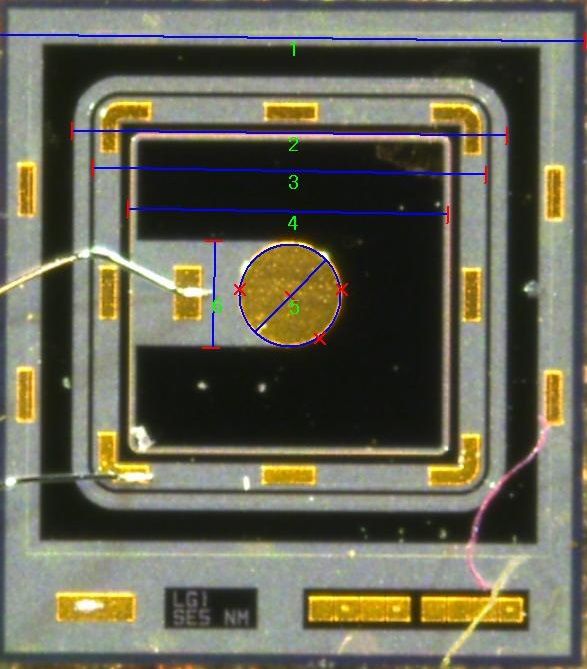}
        \caption{}
\end{subfigure}
\centering
\begin{subfigure}{0.3\textwidth}  
\includegraphics[width=\textwidth]
        {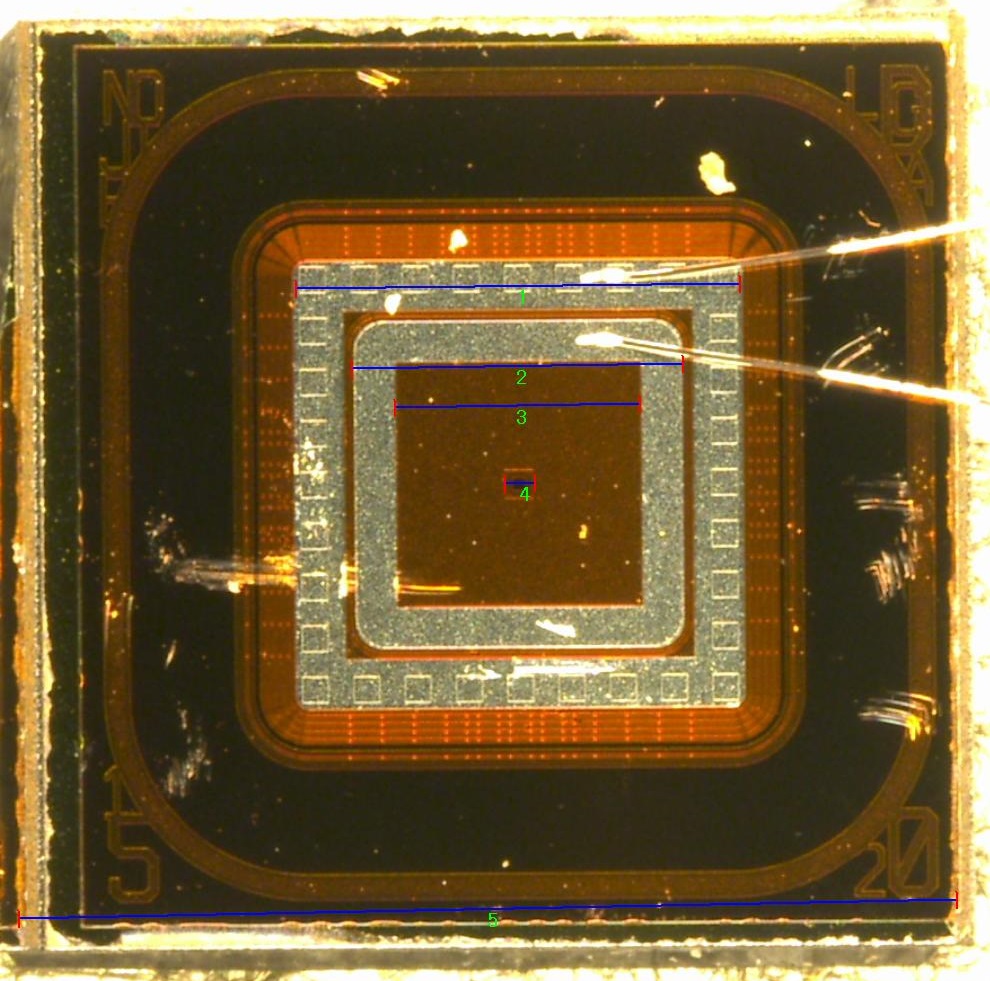}
        \caption{}
\end{subfigure}
\begin{subfigure}{0.37\textwidth}  
\includegraphics[width=\textwidth]
        {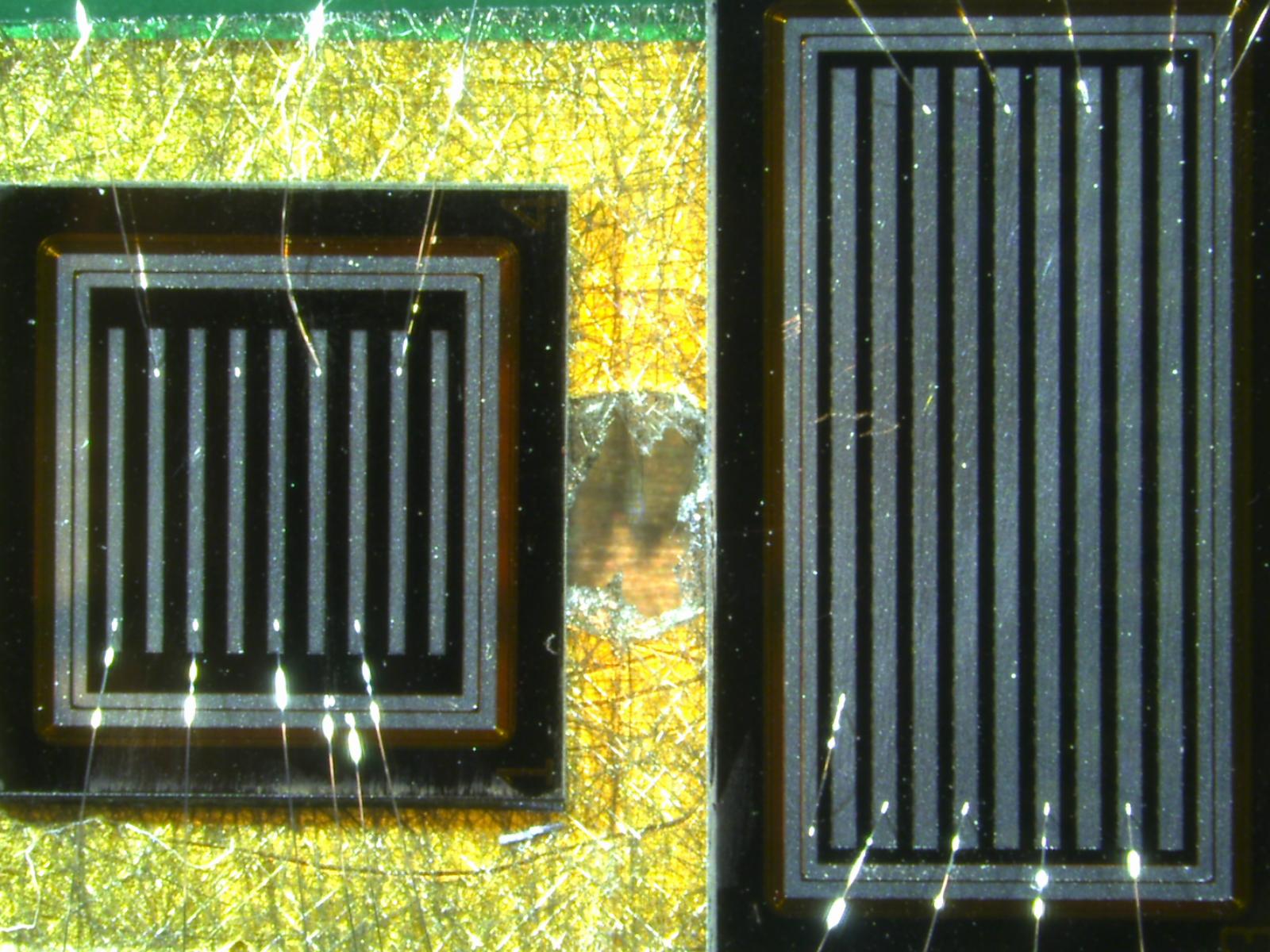}
        \caption{}
\end{subfigure}
    \caption{(a) Single pad geometry of HPK 3.1, 3.2 LGADs and PIN. (b) single pad BNL \SI{20}{\micro\meter} LGAD. (c) BNL strip AC-LGADs.}
    \label{fig:LGADS}
\end{figure}

Three standard single pads LGADs from HPK (Hamamatsu Photonics), full characteristics shown in~\cite{Padilla_2020}, and BNL (Brookhaven National Laboratories) were tested. The devices tested had \SI{50}{\micro\meter} and \SI{20}{\micro\meter} of active thickness.
A complete list of the devices is shown in Tab.~\ref{tab:LGADs}, and pictures are shown in Fig.~\ref{fig:LGADS}.
A shallow and deep gain layer in the table means a gain layer with peaking doping within \SI{1}{\micro\meter} and over \SI{2}{\micro\meter} from the top surface of the sensor, respectively.
The sensor with a deeper gain layer has an increased gain due to the effects discussed in literature~\cite{Padilla_2020}.
Unfortunately, it is not possible to estimate the gain and time resolution of the AC-LGAD prototype in the same way it was done with the other four devices. 
The charge-sharing mechanism doesn’t allow characterization using a simple beta source setup without position information. 
Studies done on similar devices at the Fermilab test beam facility, where a tracker is present, are shown in~\cite{Heller:2022aug}.

\begin{figure}[H]
    \centering
    \begin{subfigure}{0.49\textwidth}  
    \includegraphics[width=\textwidth]
        {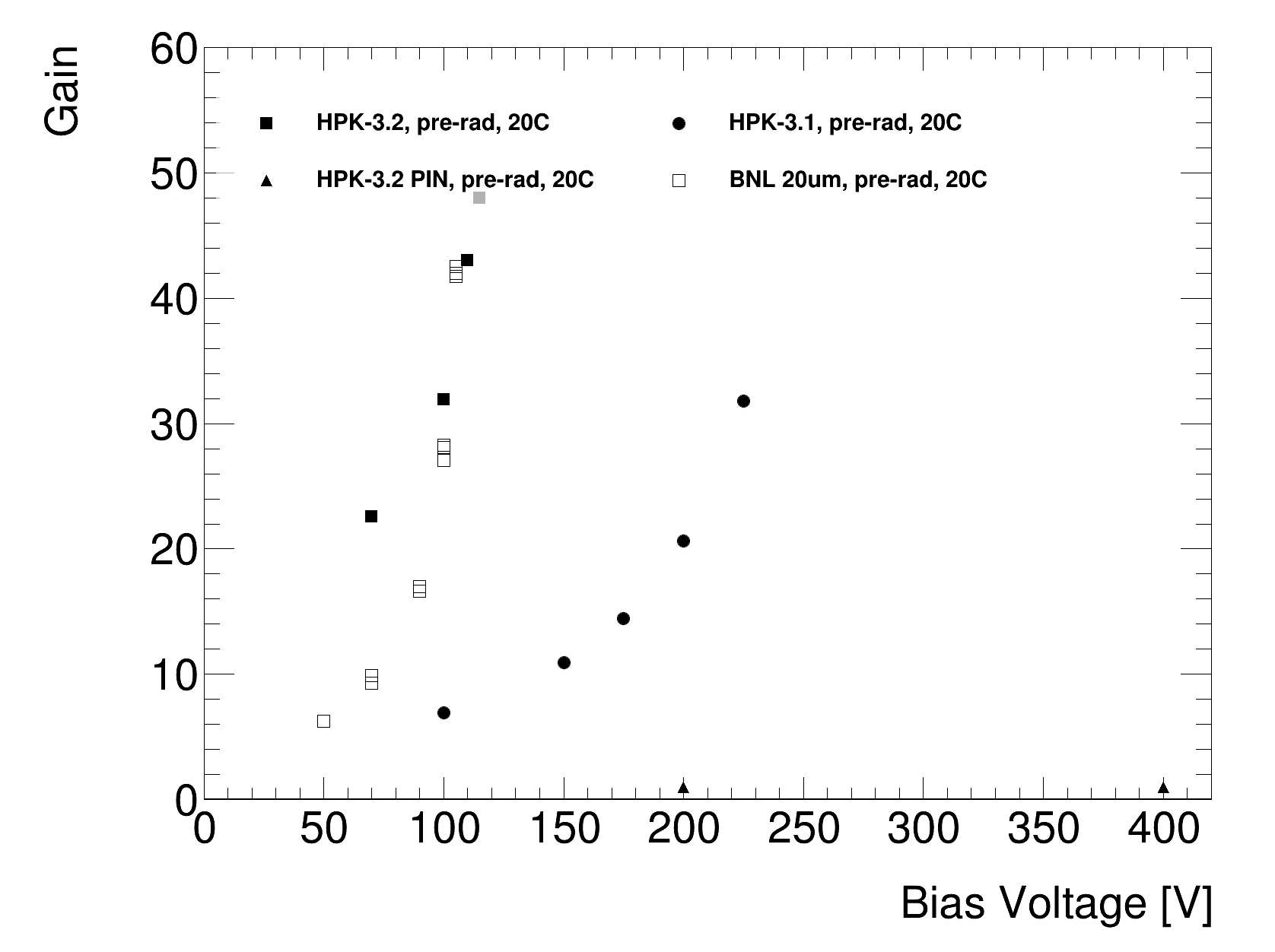}
        \caption{}
    \end{subfigure}
    \begin{subfigure}{0.49\textwidth}  
    \includegraphics[width=\textwidth]
        {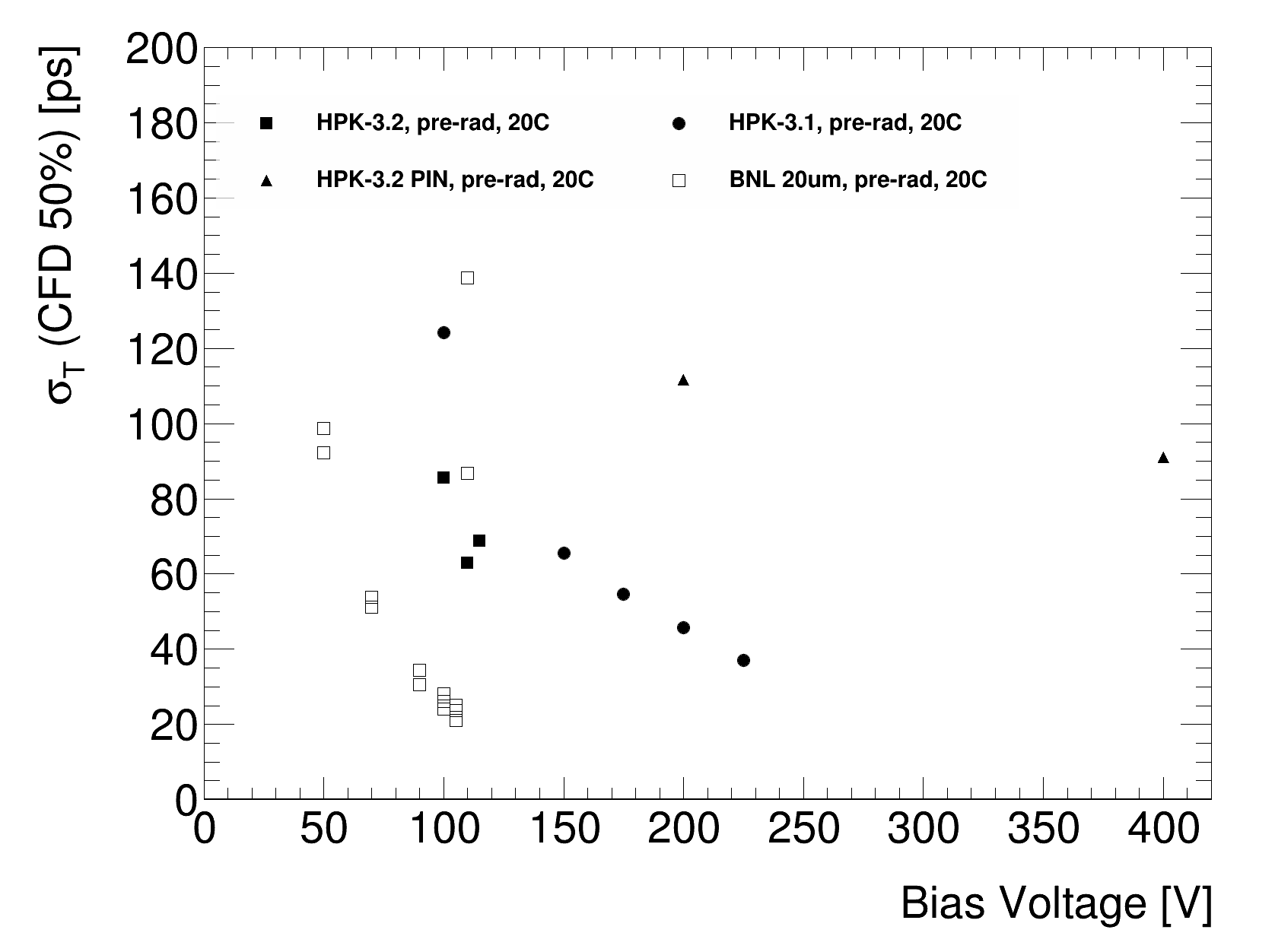}
        \caption{}
    \end{subfigure}
    \caption{(a) Gain of the devices as a function of bias voltage. (b) Time resolution of the devices for charged particles as a function of bias voltage. The values were measured with an $^{90}$Sr $\beta$ source in coincidence with a timing trigger~\cite{bib:HPKirradiation35vs50}.}
    \label{fig:LGADS_beta}
\end{figure}

All devices were tested at several voltages up to slightly less than the breakdown voltage (BV).
The gain (a) and time resolution (b) as a function of bias voltage in response to minimum ionizing particles (MIPs) for the four devices tested are shown in Fig.~\ref{fig:LGADS_beta}.
The tested LGAD, a device under test (DUT), is mounted on a fast analog electronic board (up to 2~GHz bandwidth) digitized by a GHz bandwidth digital scope.
A trigger sensor (a second HPK LGAD with a time resolution of 17ps), which acts as a time reference, is also mounted on a fast electronic board.
The electronic boards are mounted on a frame that aligns DUT and trigger to a $^{90}$Sr beta source.
A digital oscilloscope records the entire waveform of both trigger and DUT in each event, so the complete event information is available for offline analysis.
From the waveforms, the mean collected charge is calculated for each voltage and divided by the collected charge in a PIN sensor with the same thickness to calculate the gain.
The time resolution is the standard deviation of the 50\% CFD time difference between the trigger and DUT.

\begin{table}[H]
    \centering
    \begin{tabular}{|l|c|c|c|c|c|}
        \hline
         Device & Producer & BV & Thickness & Gain layer & Geometry \\
         \hline
         HPK 3.1 & HPK & \SI{230}{\volt} & \SI{50}{\micro\meter} & shallow & 1.3x1.3~\si{\milli\meter\squared}\\
         HPK 3.2 & HPK & \SI{130}{\volt} & \SI{50}{\micro\meter} & deep & 1.3x1.3~\si{\milli\meter\squared}\\
         HPK PIN & HPK & \SI{400}{\volt} & \SI{50}{\micro\meter} & no gain & 1.3x1.3~\si{\milli\meter\squared}\\
         \hline
         BNL 20um & BNL & \SI{100}{\volt} & \SI{20}{\micro\meter} & shallow & 1.3x1.3~\si{\milli\meter\squared}\\
         BNL AC-LGAD 10\si{\milli\meter} & BNL & \SI{250}{\volt} & \SI{50}{\micro\meter} &shallow & 5x10~\si{\milli\meter\squared} \\
         BNL AC-LGAD 5\si{\milli\meter} & BNL & \SI{250}{\volt} & \SI{50}{\micro\meter} &shallow & 5x5~\si{\milli\meter\squared} \\
         \hline
    \end{tabular}
    \caption{List of tested HPK and BNL LGADs.}
    \label{tab:LGADs}
\end{table}

\section{Experimental setup at SSRL}
\label{sec:setup}
\begin{figure}[H]
    \centering
    \begin{subfigure}{0.437\textwidth}  
    \includegraphics[width=\textwidth]
        {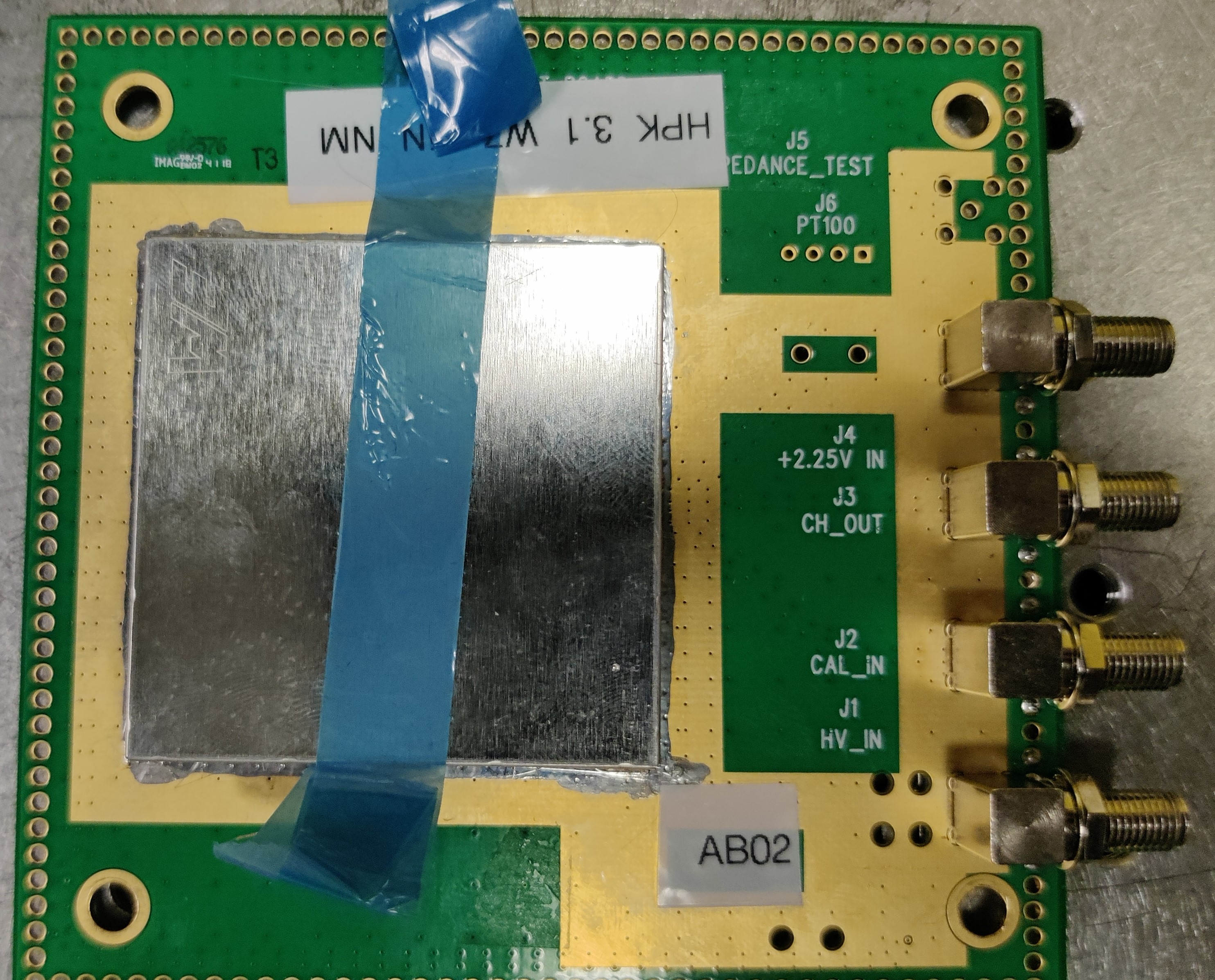}
        \caption{}
    \end{subfigure}
    \begin{subfigure}{0.47\textwidth}  
    \includegraphics[width=\textwidth]
        {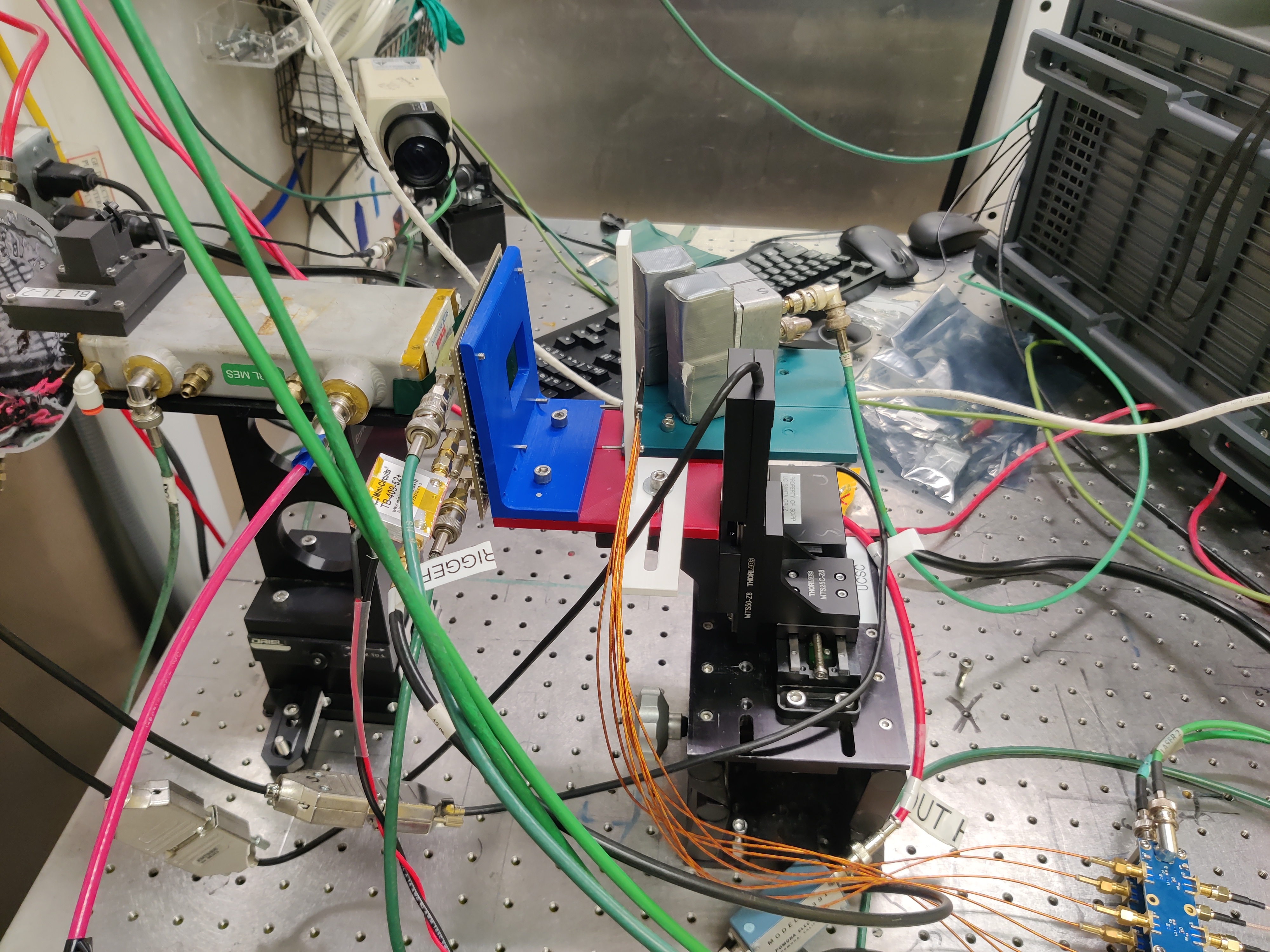}
        \caption{}
    \end{subfigure}
    \caption{(a) Single channel analog amplifier readout board used for sensor testing. (b) Data-taking setup at SSRL. The beam comes from the pipe on the left side. The grey box left of the blue holder is the gas ionization chamber to monitor the beam intensity. The readout board sits on the blue holder connected to the X-Y linear stages. }
    \label{fig:setup}
\end{figure}

The tested LGAD and PIN sensors are mounted on fast analog TIA amplifier (current mode) boards (bandwidth $\approx$ \SI{2}{\giga\hertz}), shown in Fig.~\ref{fig:setup}(a), and digitized by a fast oscilloscope or digitizer. 
The experimental setup in the beamline is shown in Fig.~\ref{fig:setup}(b). 
The fast boards have either 1 (schematics in~\cite{seiden2021potential}) or 16 channels, and the trans-impedance of the amplifiers is \SI{470}{\ohm} (plus an in-line second-stage amplifier with gain 10) and \SI{5300}{\ohm}, respectively.
The digitizer devices used are: a \SI{13}{\giga\hertz} bandwidth \SI{128}{\giga S \per \second}\ oscilloscope\footnote{Keysight Infiniium UXR} (for the single channel boards) 
and a \SI{500}{\mega\hertz}, \SI{5}{\giga S \per \second} 16-channel digitizer\footnote{CAEN DT5742} (for the 16 channel board).
The board sits on a 3D-printed frame mounted on X-Y linear stages\footnote{Thorlabs} for fine adjustment. 
In front of the board is a gas ionization chamber from the SSRL beamline instrumentation that measures the beam intensity. 
The output of the ionization chamber allows adjusting the monochromator to select the beam's baseline and harmonic (twice the nominal energy) components.
The high voltage (HV) for the sensor bias and low voltage (LV) for powering the board are provided by a  tabletop HV power supply\footnote{CAEN} and a low noise laboratory DC power supply, respectively.
The components (HV supply, motors, oscilloscope and digitizer) are remotely controlled from outside the hutch with a laptop that also controls the data taking.

\begin{figure}
    \centering
    \includegraphics[width=\linewidth]{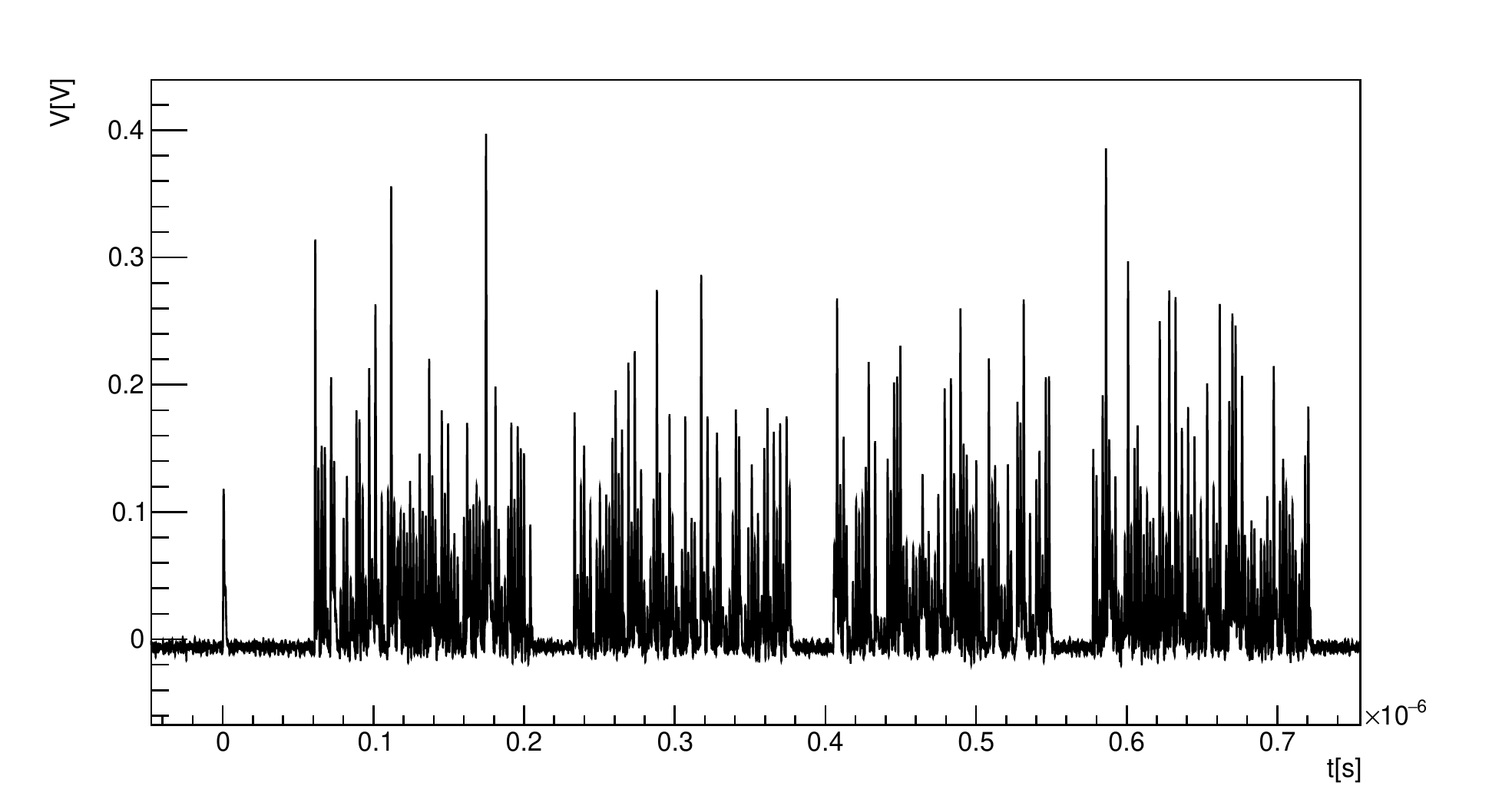}
    \caption{Oscilloscope recorded waveform from an LGAD showing the bunch structure of the SSRL. Four groups of 70 bunches in the beam are separated by an empty interval of a few tens of \SI{}{\nano\second}. The bunch structure is periodic and in phase with the \SI{1.28}{\mega\hertz} synchronizing signal used later in this work to trigger the acquisition.}
    \label{fig:bunch-structure}
\end{figure}

SSRL provides a \SI{1.28}{\mega\hertz} fast synchronization signal in phase with the cyclotron. 
This signal triggered the data acquisition, allowing each event to have the same bunch structure in time.
The beam structure has a window of 70 bunches spaced in time by \SI{2.1}{\nano\second} separated by a few tens of \SI{}{\nano\second} intervals without buckets, which are crucial to evaluate the baseline for each event. 
An example of the recorded signal from an LGAD showing the beam structure is presented in Fig.~\ref{fig:bunch-structure}. At time zero, there's a single separated synchronization bucket.
The single pulses are not visible in the figure because the time scale is too large, but pulses with  \SI{2.1}{\nano\second} time distance are fully separated, as seen in Fig.~\ref{fig:waveforms}.

\section{Geant4 simulation}
\label{sec:geant4}
\begin{figure}
\centering
    \includegraphics[width=\textwidth]{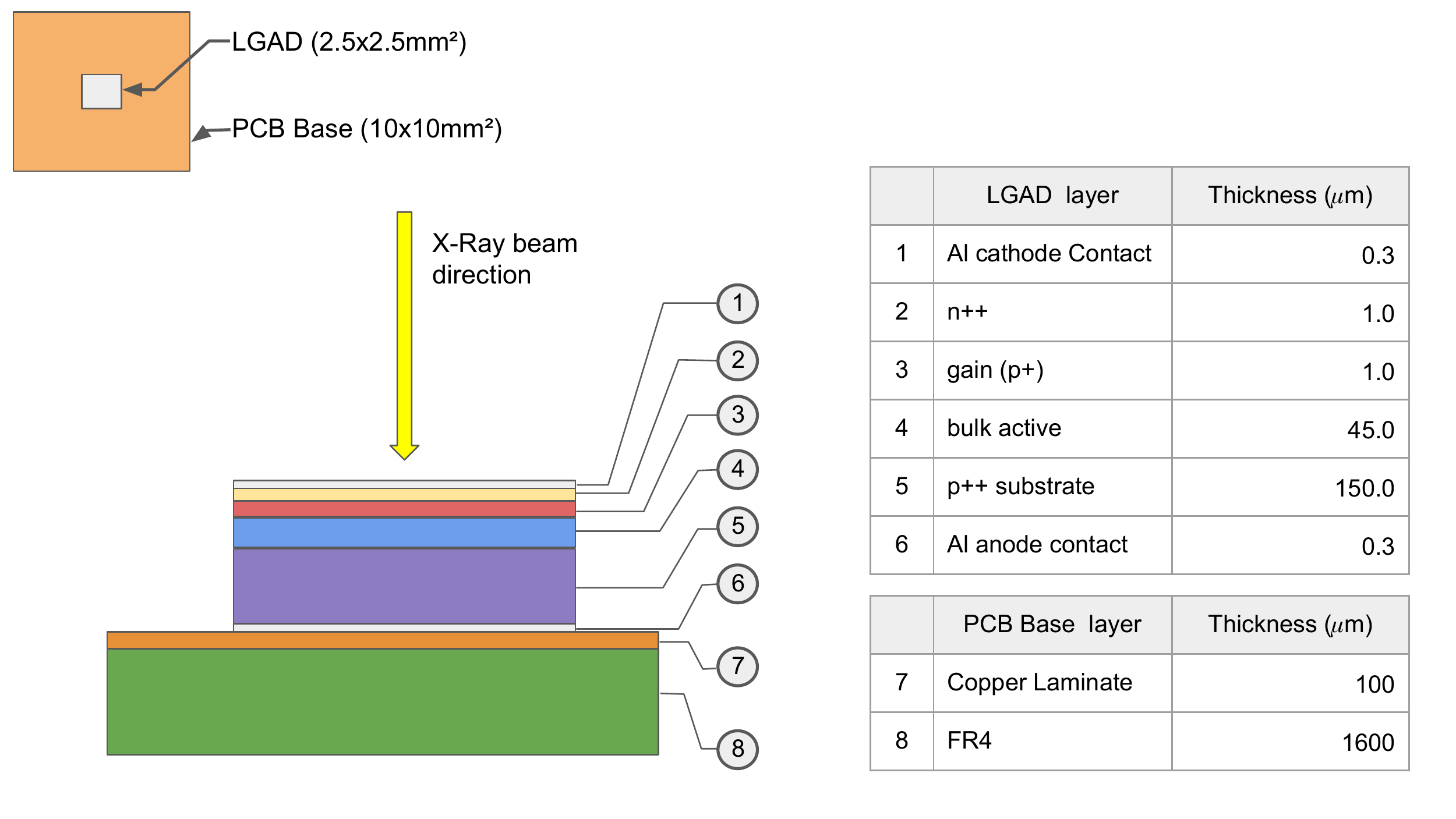}
    \caption{An schematic view of geometry (not in scale) of a HPK type 3.2 LGAD and support structure used in the Geant4 simulation. The direction of the beam is indicated in the figure, as well as the relevant structure cross section dimensions and materials.}
    \label{fig:g4_geometry}
\end{figure}

A simulation of the interaction of X-rays with the test setup was performed using Geant4 Toolkit\footnote{Version 11.1.1} ~\cite{Allison_1610988,ALLISON2016186,AGOSTINELLI2003250}. The simulated geometry was comprised of a HPK type 3.2 LGAD sensor attached to a printed circuit board (PCB), acting as a support structure. The  geometry and materials for the simulation were described in GDML~\cite{Chytracek_1710291} . Fig.~\ref{fig:g4_geometry} shows all relevant dimensions of the sensor and support as entered in the simulation. To replicate the test beam conditions, the structure was surrounded by air at room temperature.  

The simulation uses the Geant4 standard EM Physics List, and the particles where tracked using production cuts of \SI{100}{\nano\meter} and \SI{100}{\electronvolt}. All simulations were performed for 1M photons with energies between 5 to 70 \si{\kilo\electronvolt}, in 5 \si{\kilo\electronvolt} steps. No spread in the nominal energy was included in the simulation since the beam energy spread at the SSRL is negligible ($\Delta E/E \approx 10^{-4}$) \cite{SSRL112}. 


\subsection{Simulation of the energy deposition}

One of the main objectives of the simulation is to understand the energy deposition in the different LGAD layers. Differently from the case of high energy physics applications, where a minimum ionizing particle produces ionization charges along its track across the sensor structures, in X-rays applications the charge will be concentrated at the position where the photon is absorbed, with a probability  that will depend on the photon energy.  

\begin{figure}[H]
\centering

\includegraphics[width=\textwidth]
        {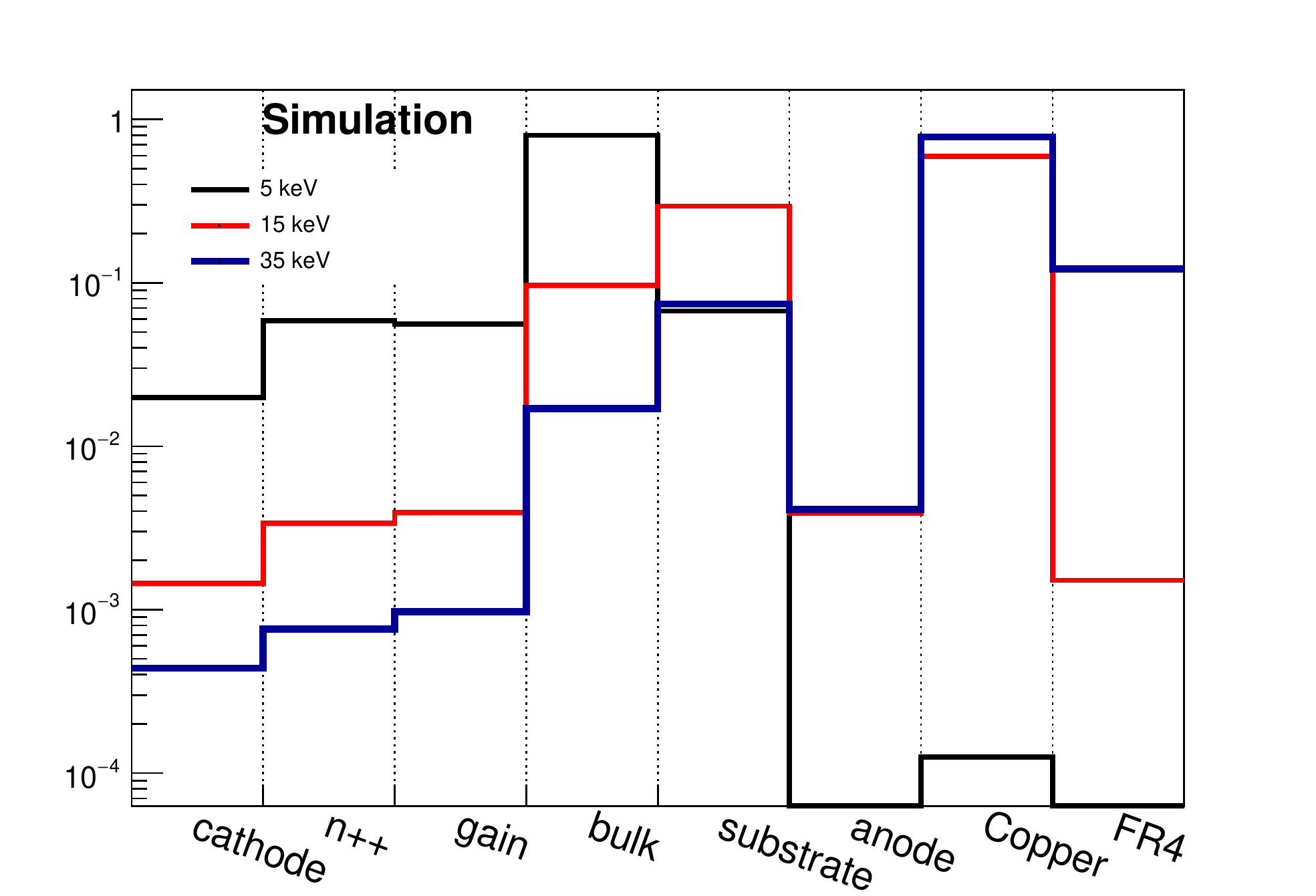}

\caption{Fraction of the number of conversions occurring at each region of the experimental setup for  X-rays energies of 5, 15 and 35 \si{\kilo\electronvolt} (refer to Fig.~\ref{fig:g4_geometry} for a region description).}
\label{fig:volumes}
\end{figure}

To simulate the number of primaries created in each layer of the sensor and support PCB, the X-ray beam was changed in the simulation to a point-like geometry hitting the center of the LGAD structure in the perpendicular direction. Fig.~\ref{fig:volumes} shows the fraction of primary charges produced in each layer of the setup for $10^6$ X-ray photons with energies of 5, 15, and 35 \si{\kilo\electronvolt}. This simulation can provide relevant guidance on the sensor fabrication, as only the charges produced inside the sensitive area of the sensor (gain and bulk layers) contribute to the signal generation. 
The only practical way to increase the efficiency of the sensor is to increase the bulk depth. We also want to minimize the conversions in the gain layer, as the intense electrical field will cause large fluctuations in the final charge produced in the device. Moreover, it is well known that thinner sensors provide the best timing~\cite{bib:UFSD300umTB}. The trade-off between the detection efficiency and timing should be evaluated together with a full TCAD simulation of the sensor.

\section{TCAD simulation}
\label{sec:TCAD}
TCAD software from Synopsys Sentaurus~\cite{sentaurus} is used to simulate the LGAD response to X-ray photons. Only photons that are fully absorbed are simulated to produce the results shown in this section. Compton interaction was not considered in the TCAD studies. The photon absorption is modeled with point-like energy transfer that corresponds to the photon energy, and the generated electron-hole pairs are distributed in a \SI{1}{\micro\metre^2} box. 
The simulated current signal at the device level is convoluted with a transimpedance amplifier~\cite{TIA} (TIA) SPICE~\cite{SPICE} model to match the experiment readout~\cite{GALLOWAY20195}, and a software digitization rate of \SI{128}{\giga S \per \second} is applied to replicate the oscilloscope sampling.


\subsection{Time variation due to different absorption depths}

\begin{figure}[H]
    \centering
    \begin{subfigure}[t]{0.48\textwidth}   
        \includegraphics[width=1.0\linewidth]{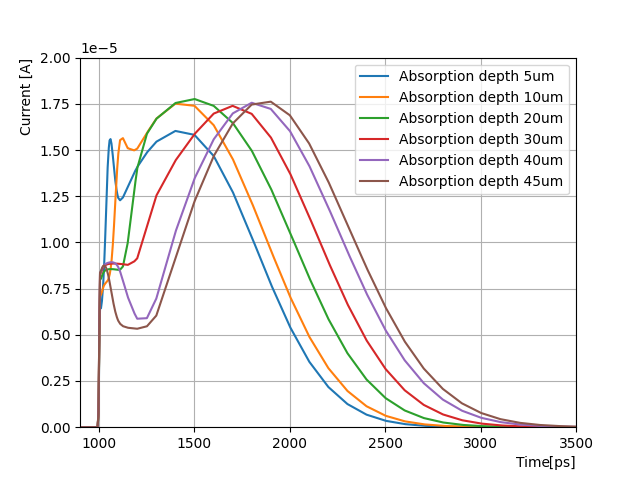}
        \caption{Current signal at the device level.}
        \label{fig:tcad:tcad_depth_signal}
    \end{subfigure}
    \begin{subfigure}[t]{0.48\textwidth}   
        \includegraphics[width=1.0\linewidth]{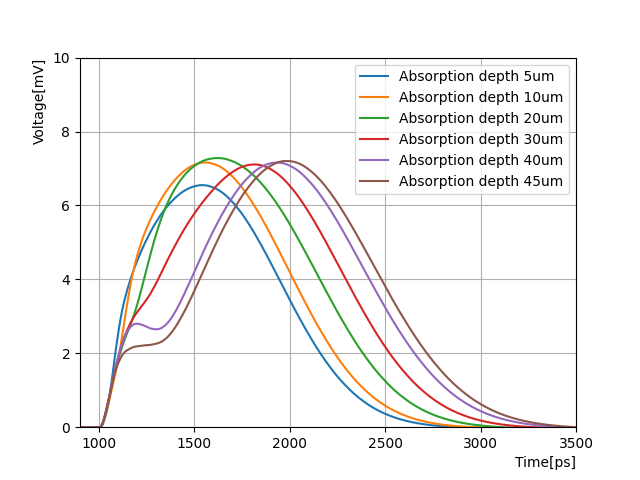}
        \caption{Convoluted voltage signal with TIA.}
        \label{fig:tcad:amp_depth_signal}
    \end{subfigure}
    \caption{\SI{50}{\micro\meter} LGAD signal response to a single \SI{20}{\keV} photon at different absorption depth.}
    \label{fig:tcad:depth_signal}
\end{figure}

Significant ($>$\SI{100}{\pico\second}) variation in the time of arrival of the signal is observed in the measurement, and a potential explanation for this is related to the depth of the photon absorption. 
To study this effect, single photon absorptions with an energy of \SI{20}{\keV} at different depths are simulated. 
The current signals at the device level and convoluted voltage signals are shown in Fig.~\ref{fig:tcad:depth_signal}. 
The time variation on the rising edge is due to the drifting time for generated charges to arrive at the multiplication layer. Multiplication for charge generated at the bottom of the device will have a delay of \SI{500}{\pico\second}, which corresponds to the travel time of the electrons to reach the gain layer for a bulk thickness of \SI{50}{\micro\metre}. 
As seen in Fig.~\ref{fig:tcad:depth_signal}(b), the variation can be reduced by using the initial instantaneous current raise. 
The time of arrival to evaluate the time resolution will therefore be calculated in the following section using the 20\% of the pulse (constant fraction discriminator method, 20\% CFD). 

\subsection{Multiple photons conversion and harmonic energy}

\begin{figure}[H]
    \centering
    \includegraphics[width=0.7\linewidth]{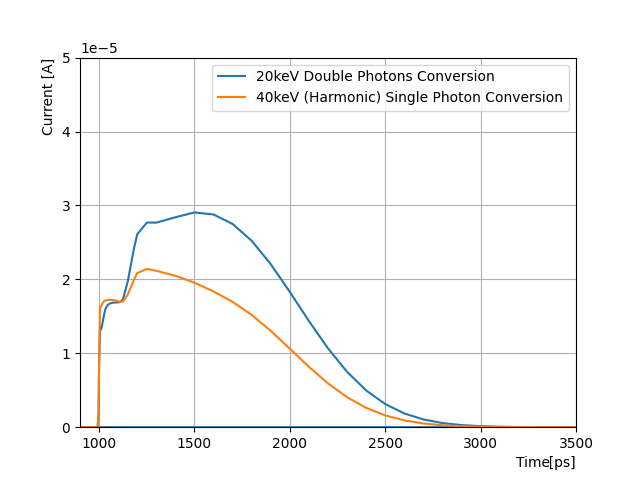}
    \caption{Comparison of the generated signal of double photon conversion (blue) vs single photon from the harmonic (twice the energy) components (orange).}
    \label{fig:tcad:multi_photon}
\end{figure}

Multiple photons conversions within the device can generate the same amount of  electron-hole (e-h) pairs corresponding to the harmonic energy components. For example, the amount of generated e-h pairs corresponds to the total deposited energy of \SI{40}{\keV} can originate from either two photons conversion with the energy of \SI{20}{\keV} each, or a single photon of \SI{40}{\keV} from the beam harmonics with the energy of \SI{40}{\keV}. 
In Fig.~\ref{fig:tcad:multi_photon}, there's a significant difference in the signal response due to double photon conversion vs. single photon from the harmonic. The gain suppression mechanism due to electric field reduction from the high density of e-h pairs in the gain layer can explain this difference. Effectively, the device's gain in the case of the \SI{40}{\keV} X-ray absorption is less than in the case of two \SI{20}{\keV} absorption because of the different charge density in the interaction point.
Extensive TCAD simulations of the effect were presented in~\cite{TCAD_saturation}.

\subsection{Gain dependence on shape of energy deposition}
\label{sec:TCAD_sat}
\begin{figure}[H]
    \centering
    \includegraphics[width=0.65\linewidth]{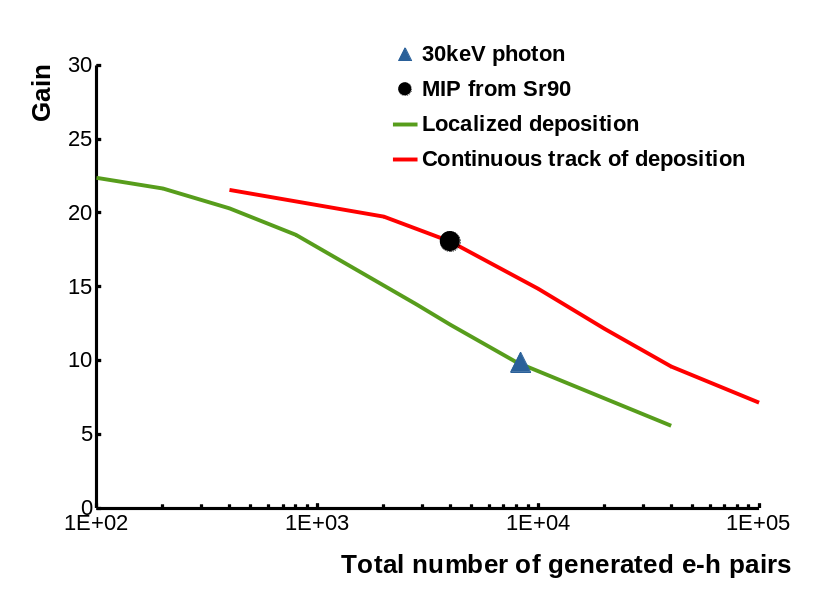}
    \caption{TCAD simulated gain for continuous track of energy deposition (MIP) and localized energy deposition (X-ray). The gain corresponding to a MIP from $^{90}$Sr and a 30~keV X-ray are highlighted in the plot.}
    \label{fig:tcad:track_vs_local}
\end{figure}

The gain dependence on the energy deposition characteristics inside the device was studied with the aid of TCAD simulation.
The simulated gain from the track of deposited energy (MIP-like) is compared to localized deposition (X-ray-like). 
In the model, the deposited energy for the track is evenly distributed along \SI{50}{\micro\meter} with a width of \SI{0.5}{\micro\meter} inside the sensor, and the energy of localized deposition is generated within a \SI{1}{\micro\meter^3} box. 
The gain for these two scenarios is shown in Fig.~\ref{fig:tcad:track_vs_local} for a \SI{50}{\micro\meter} LGAD with gain 23 and saturated drift velocity. 
The track deposition generally has a higher gain than the localized deposition for the same number of generated e-h pairs. 
This TCAD simulated result is consistent with the observation that the gain for X-ray photons is lower than the measurement with MIPs shown in the following sections.
This behavior is again a result of the gain saturation effect in LGADs~\cite{TCAD_saturation}.

\section{Average pulse analysis}
\label{sec:pulses}
\begin{figure}
    \centering
    \subfloat[]{ \includegraphics[width=0.49\linewidth]{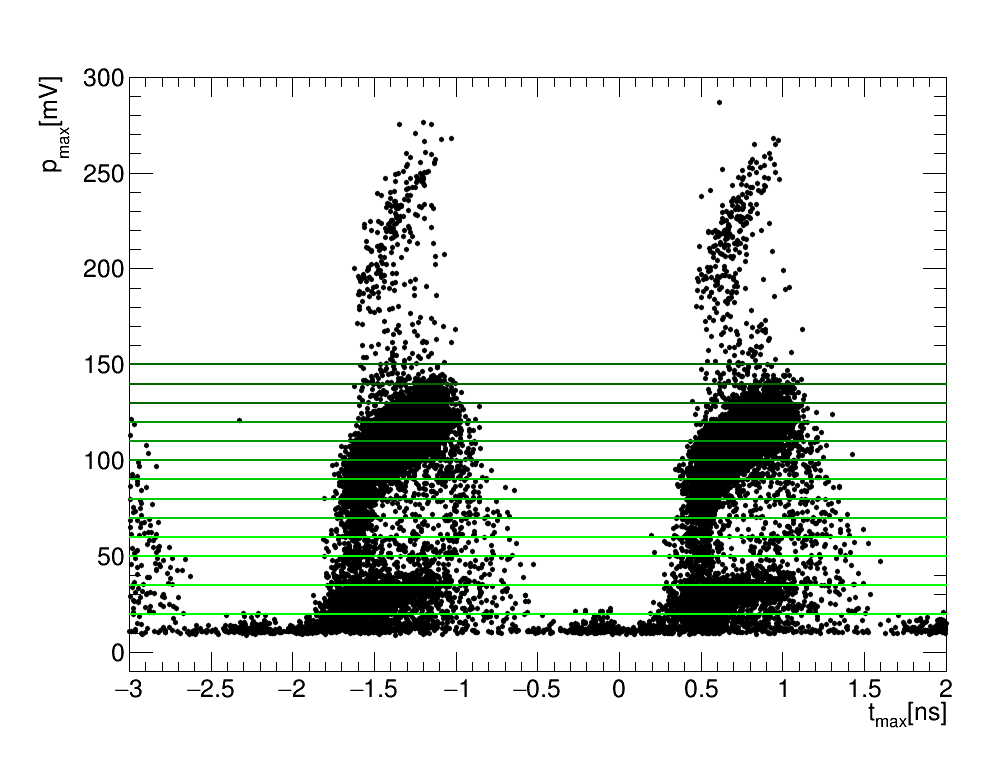} }
    \subfloat[]{ \includegraphics[width=0.49\linewidth]{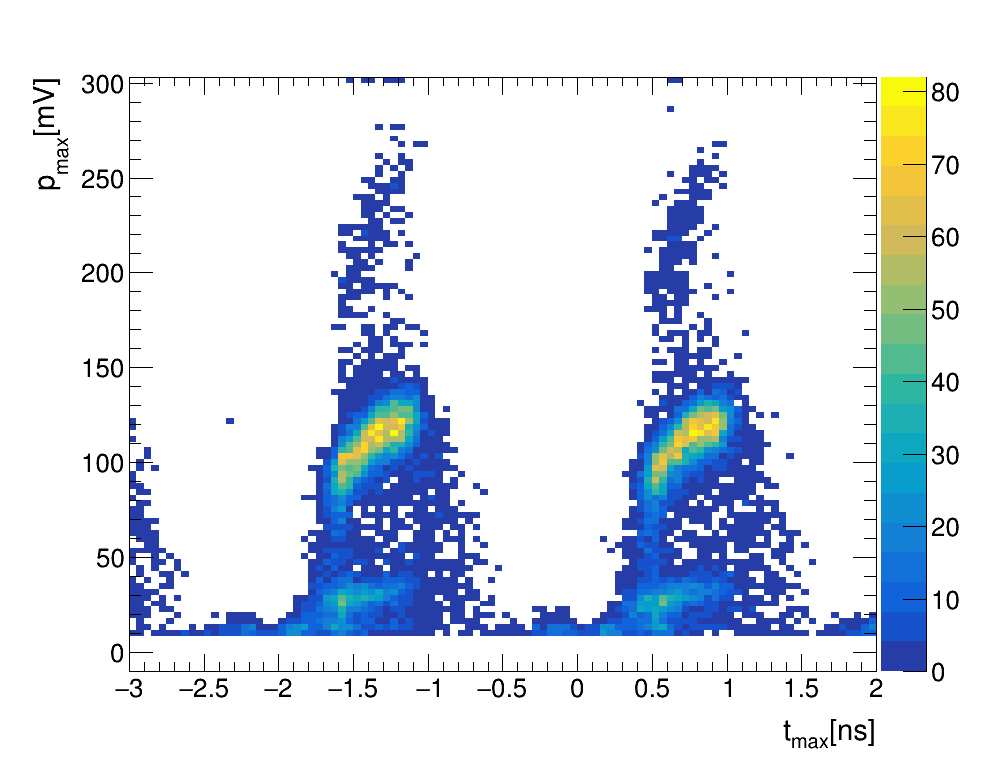} } \\
    \caption{(a) Plot of $\pmax$ (pulse maximum) vs $\tmax$ (time of the pulse maximum) for HPK~3.1 sensor biased at \SI{200}{\volt} and X-ray energy of \SI{30}{\kilo\electronvolt}. Colored lines correspond to sub-selections used to draw the averaged pulse shapes in Fig.~\ref{fig:pulses}. (b) Same plot as in (a) but with color indicating the number of events in a given bin of $p_{max}$ and $t_{tmax}$.}
    \label{fig:tmaxpmax}
\end{figure}

The distribution of the pulse maximum amplitude ($\pmax$) and time of the pulse maximum ($\tmax$) for each event is shown in Fig.~\ref{fig:tmaxpmax}. The plot is for sensor HPK~3.1 biased at \SI{200}{\volt} for X-rays of \SI{30}{\kilo\electronvolt}. 
A selection on $\pmax$ of at least \SI{10}{\milli\volt} was applied to remove noise events.
The distribution is repeated every \SI{2.1}{\nano\second} as expected from the SSRL cycle.
In the distribution, a few different populations of events are divided as follows.
Until \SI{50}{\milli\volt} of $\pmax$ is a group of events likely coming from Compton interaction between X-ray and Silicon bulk. 
Between \SI{50}{\milli\volt} and \SI{150}{\milli\volt} of $\pmax$ is the main group of events for \SI{30}{\kilo\electronvolt} X-rays.
The depth of X-ray absorption causes the variation in $\tmax$, as expected from the simulation (Fig.~\ref{fig:tcad:depth_signal}).
The events for $\pmax$ over \SI{150}{\milli\volt} are for either double photon absorption or from the harmonic component with an energy of \SI{60}{\kilo\electronvolt}. 
%
\begin{figure}
    \centering
        \begin{subfigure}[t]{0.49\textwidth}   
        \includegraphics[width=1.0\linewidth]{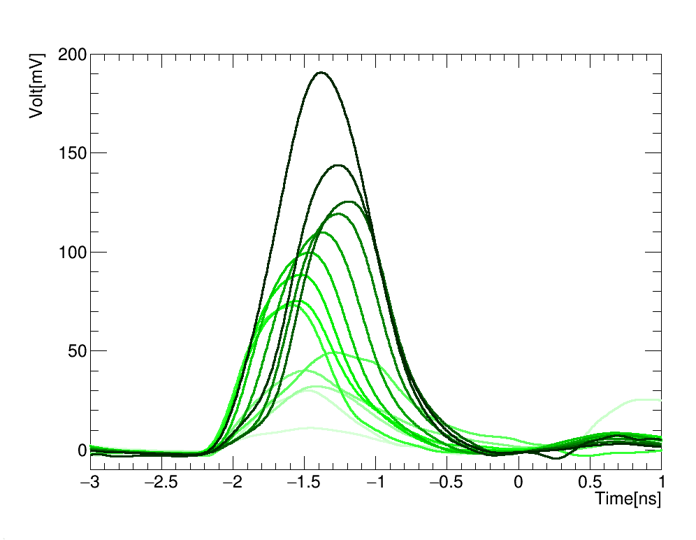}
        \caption{}
    \end{subfigure}
    \begin{subfigure}[t]{0.49\textwidth}   
        \includegraphics[width=1.0\linewidth]{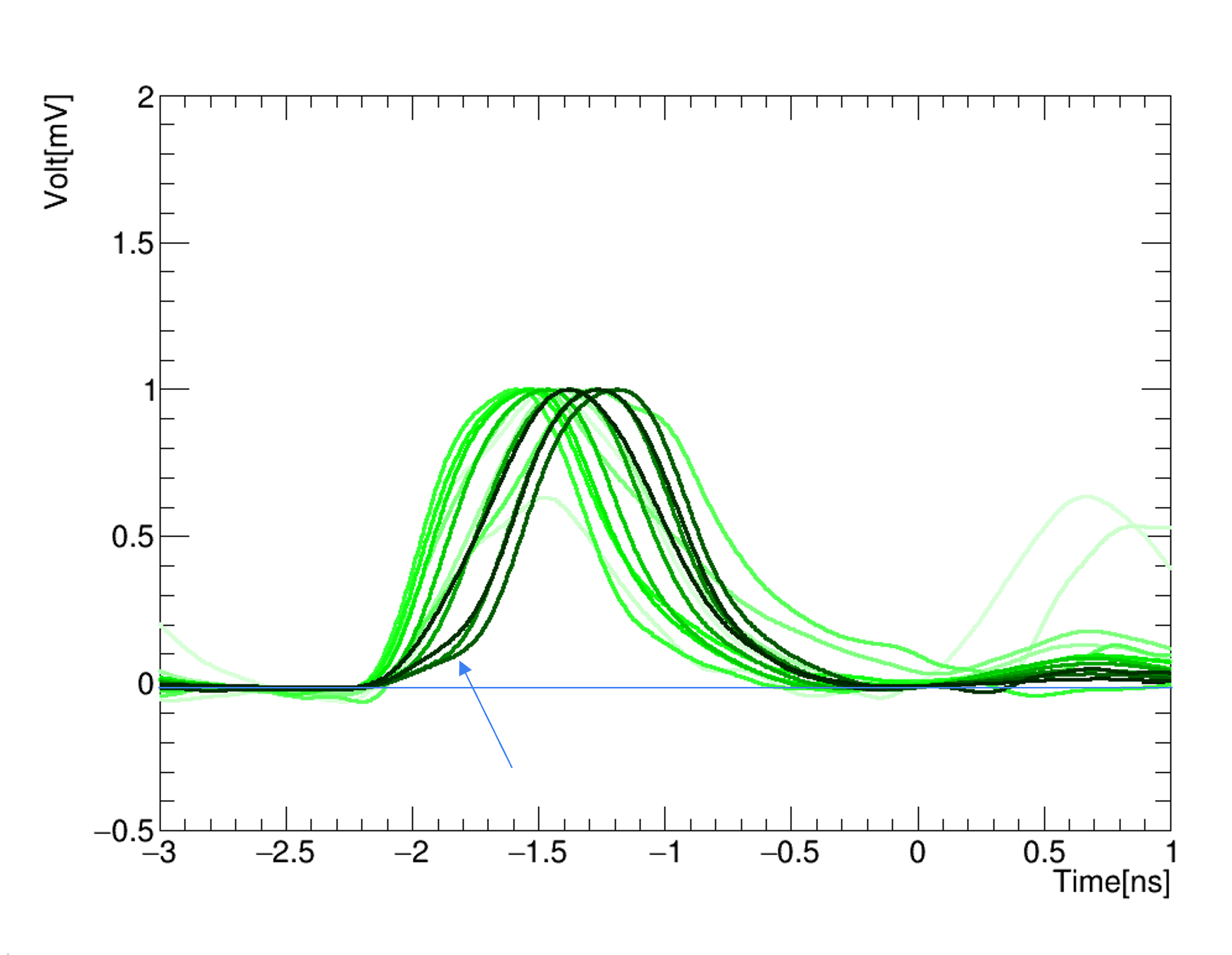}
        \caption{}
        \label{fig:pulses_normalized}
    \end{subfigure}
    \caption{(a) Averaged pulses for a sub-selection of the $\tmax$/$\pmax$ plot in Fig.~\ref{fig:tmaxpmax}. The line color in the pulses correspond to the selection lines in the $\pmax$/$\tmax$ plot. 
    (b) Same pulses but normalized. The baseline is highlighted with a blue line and the delayed pulse rise is highlighted by the blue arrow.}
    \label{fig:pulses}
\end{figure}
A slight gain loss is expected when the photon interacts close to the gain layer, as seen in Fig.~\ref{fig:tcad:depth_signal}.
This can be explained as the effect of the charge density on the electric field, which is reduced, and therefore, the gain is reduced.
When the charge is deposited deeper into the detector, the electron lateral drift reduces the charge density in the gain layer.
However, the observed direct correlation between $\tmax$ and $\pmax$ in the main distribution hints at a larger gain reduction from the one predicted in the simulation.
The effect is still observed, although reduced, at lower voltages (with lower gain) and with lower X-ray energy where the gain reduction from the charge density is less.
Fig.~\ref{fig:pulses} shows the average color-coded waveforms for event selections in $\pmax$ corresponding to the lines drawn in Fig.~\ref{fig:tmaxpmax}.
The waveforms are both un-normalized, highlighting the $\pmax$ trend, and normalized, highlighting the $\tmax$ trend.
The normalized distribution shows that the pulses with larger $\pmax$ have a delayed initial rising edge.
The initial (within the first 500~ps) step region of the pulse, predicted in Fig.~\ref{fig:tcad:depth_signal}, is observed as a different initial slope in the pulse with the highest delay indicated by the arrow.

\section{Data analysis}
\label{sec:analysis}
For each test condition (sensor type, bias and X-ray beam energy), about 15 thousand waveforms of \SI{100}{\nano\second} triggered by the SSRL synchronizing signal rising edge were digitized at \SI{128}{\giga S \per \second}. Photons from the \SI{10}{\pico\second} packet in the beam that are converted inside the LGAD active region will produce a pulse with amplitude proportional to the charges created and multiplied in the device, as it is shown in Fig.~\ref{fig:waveforms}. Because there is a significant baseline shift after each event, a baseline correction procedure was devised to remove this effect from the data. The peaks on each segment were identified by a peak finding algorithm that takes into account the electronic noise level, the peak width and the separation between neighboring peaks. The maximum digitized sample is taken as a simple estimator of the signal amplitude and the timing was estimated using a CFD method taking as a reference the information of packet separation (\SI{2.1}{\nano\second}). The following sections describe in detail each of these steps and the energy and timing resolution of the tested LGADs.

\subsection{Baseline Correction}

\begin{figure}
    \centering
    \begin{subfigure}{0.49\textwidth}   
        \includegraphics[width=\linewidth]{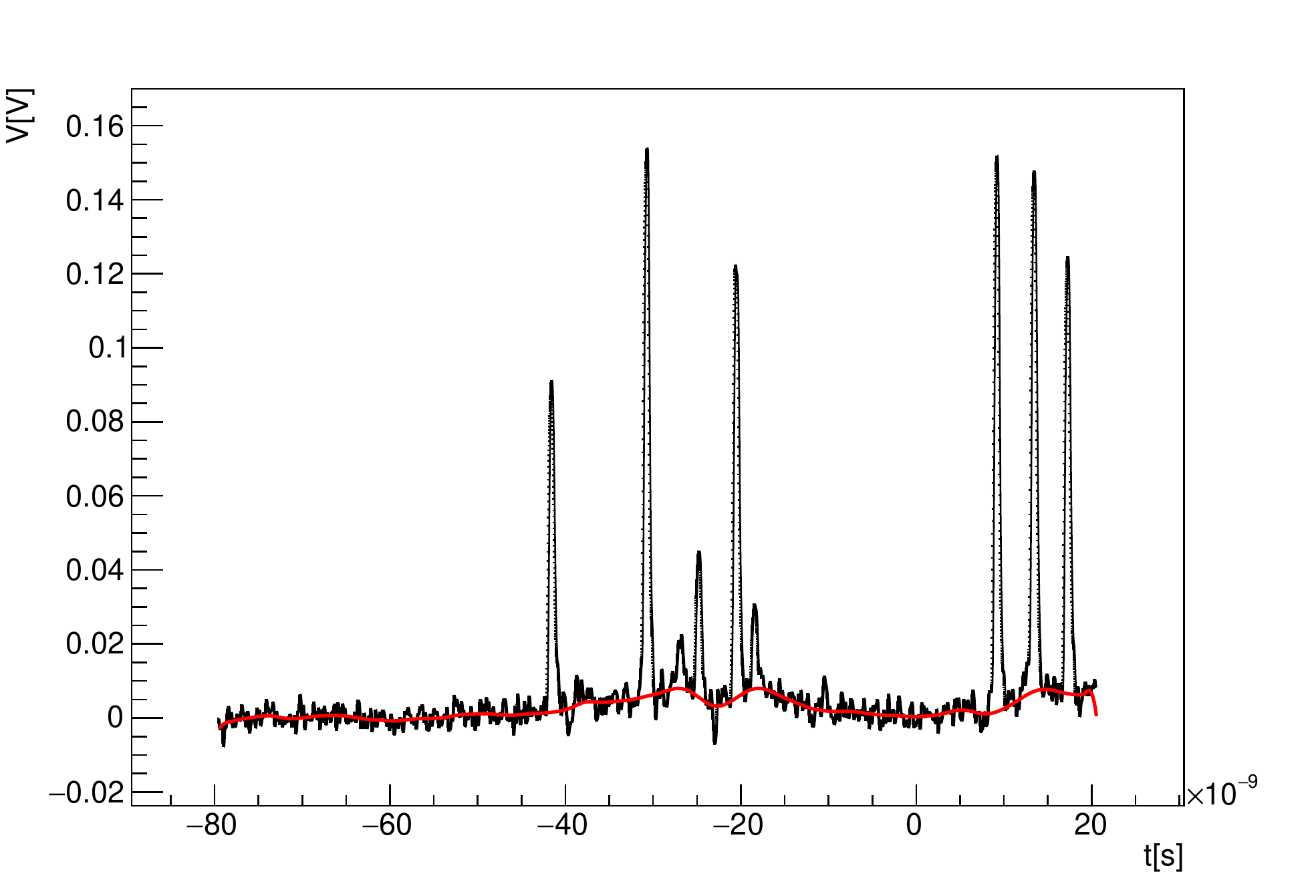}
        \caption{}
    \end{subfigure}
    \hfill  
     \begin{subfigure}{0.49\textwidth}   
        \includegraphics[width=\linewidth]{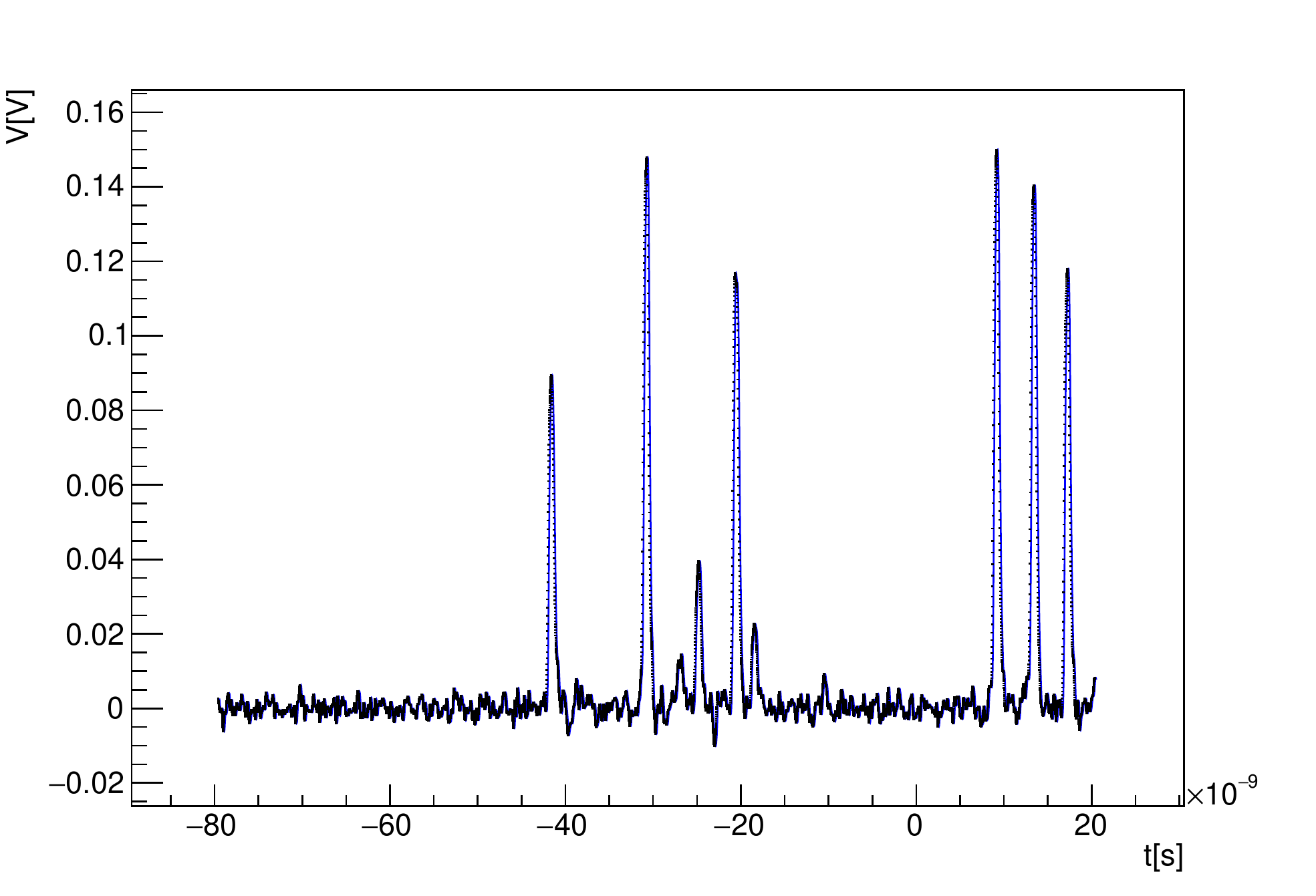}
        \caption{}
    \end{subfigure}
    \caption{(a) Waveform from a HPK Type 3.1 LGAD sensor biased at \SI{200}{\volt} exposed to a beam of \SI{35}{\kilo\electronvolt} X-rays as acquired by the oscilloscope. The baseline estimation (red curve) follows the procedure described in the text.  (b) The same waveform after baseline subtraction.}
    \label{fig:waveforms}
\end{figure}

As it can be seen in Fig.~\ref{fig:waveforms}, the baseline fluctuation in the waveform can be understood as a low-frequency signal added to the sensor response to photons (peaks). This fluctuation -- an effect introduced by the amplification circuit --  can be subtracted using an approach from spectroscopy known as {\em asymmetrically re-weighted penalized least squares smoothing} \cite{BAEK_baseline_correction}. While this method was very successful in correcting the signal baseline, it requires a significant amount of computing resources to correct the large datasets of fast sampled waveforms. However, because of the much lower frequency of the baseline fluctuation, it is possible to downsample the signal and preserve  the information needed to evaluate the  baseline shift, speeding up significantly the processing. The baseline estimated is upsampled back to the original sampling rate and subtract from the original signal, as it is shown by Fig.~\ref{fig:waveforms}.

\FloatBarrier
\subsection{Peak Finding}
\label{subsection:peak_finding}

\begin{figure}[H]
    \centering
    \begin{subfigure}{0.49\textwidth}   
        \includegraphics[width=\linewidth]{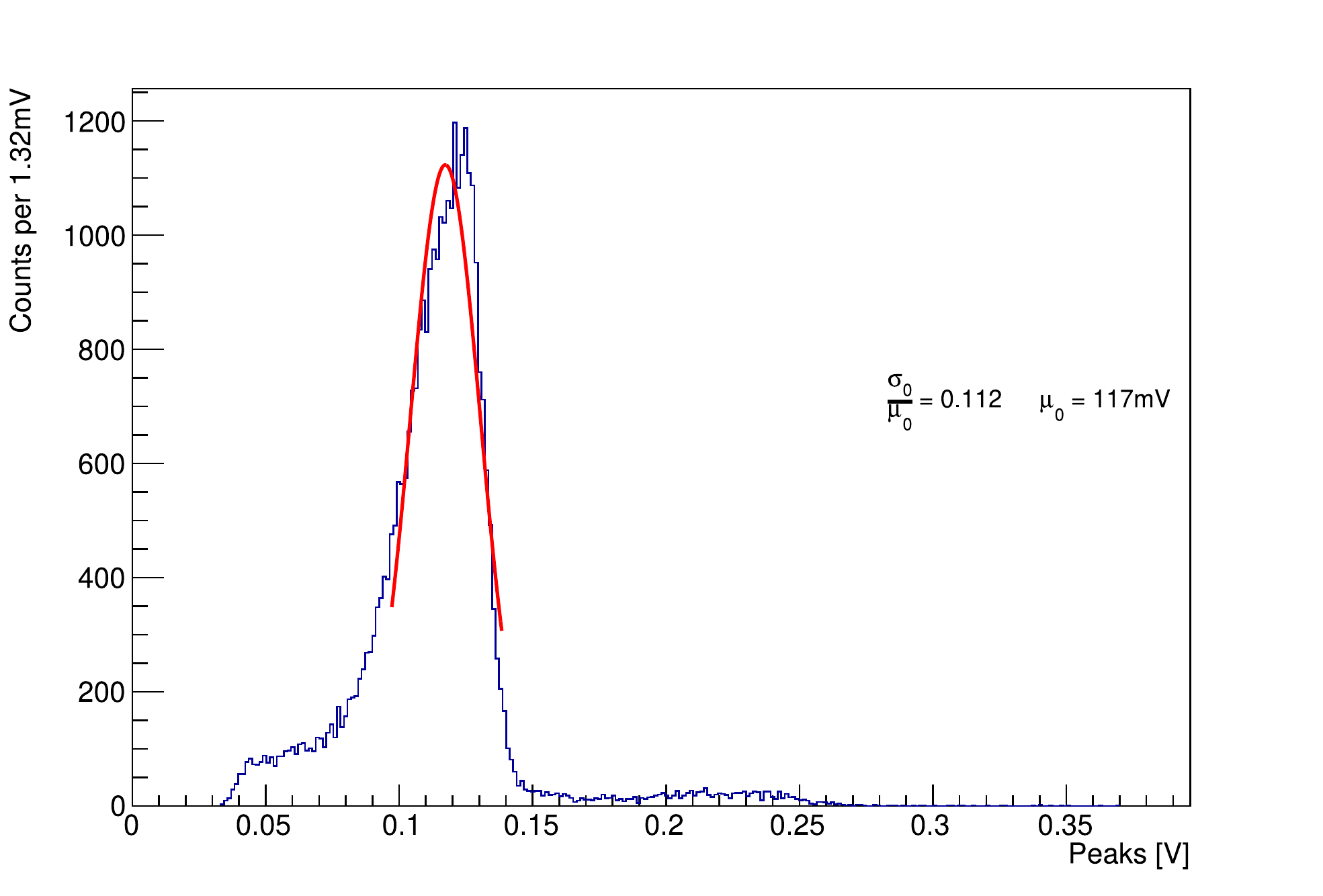}
        \caption{}
    \end{subfigure}
    \hfill  
     \begin{subfigure}{0.49\textwidth}   
        \includegraphics[width=\linewidth]{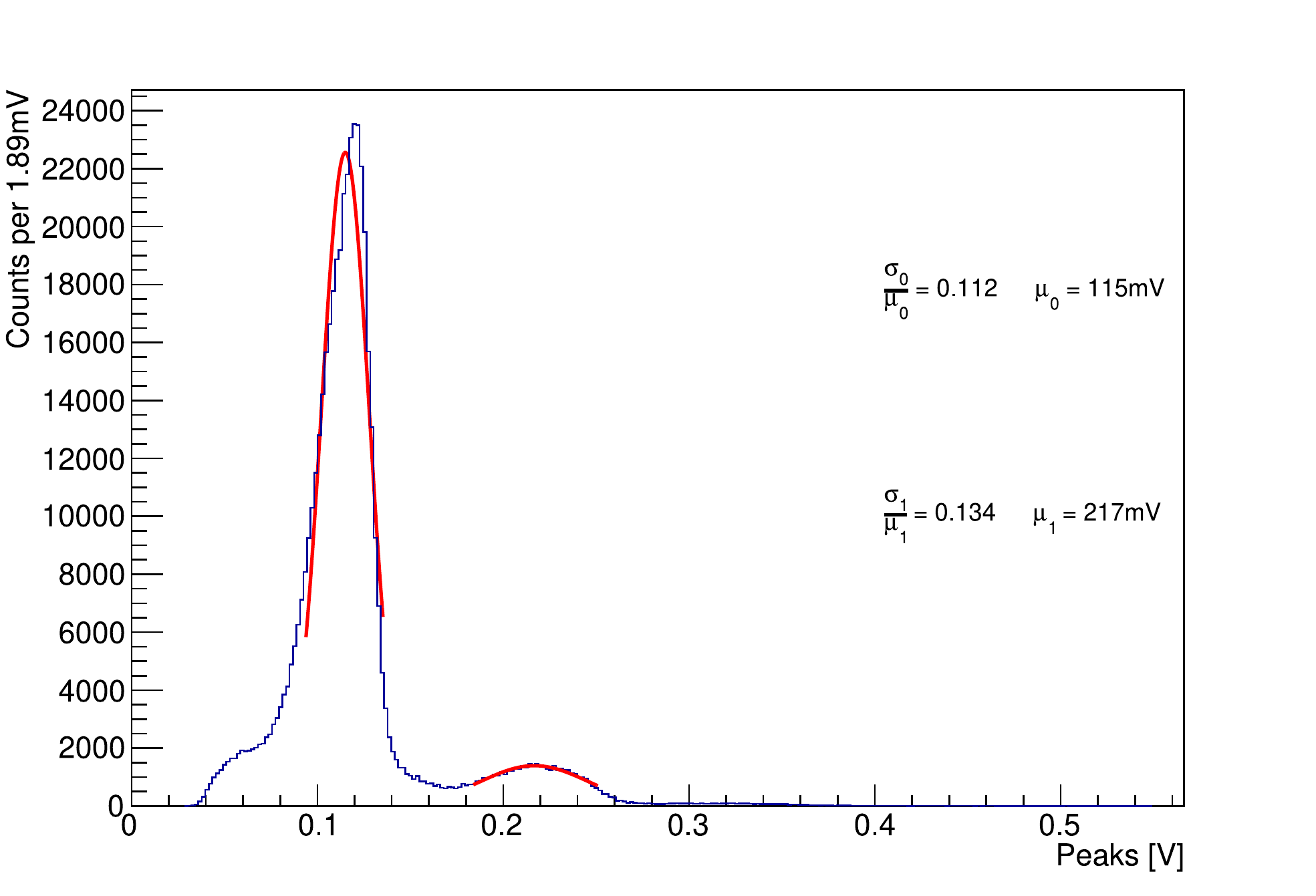}
        \caption{}
    \end{subfigure}
    \caption{(a): $\pmax$ distribution for a HPK type 3.1 LGAD sensor biased at \SI{200}{\volt} exposed to a beam of  \SI{30}{\kilo\electronvolt} X-rays with the harmonics filtered out. (b): The same setup now with the beam line monochromator adjusted to pass both the fundamental and harmonics content. The first distribution corresponds to pulses from the fundamental component of the beam and second distribution (twice the amplitude) of the harmonic. The red curves represents a Gaussian fit on the distributions.}
    \label{fig:pmax_distribution}
\end{figure}

Only the pulses in the baseline-corrected waveform above a certain noise level threshold are used to estimate the converted photon energy in the LGAD sensor. The noise was evaluated using a waveform region corresponding to an empty beam region where no X-rays packets were produced. Moreover, only signals separated in time by multiples of \SI{2.1}{\nano\second} were considered valid peaks. 
The selected pulses' maximum amplitude sample ($\pmax$) was then taken as an energy estimator. 
Fig.~\ref{fig:pmax_distribution}(a) shows the amplitude signal distribution for an HPK Type 3.1 LGAD sensor exposed to a 30 keV X-ray beam. The SSRL beamline monochromator can be adjusted to filter a given fraction of harmonics corresponding to exactly twice the nominal beam energy. 
Fig.~\ref{fig:pmax_distribution}(b) shows the pulse amplitude distribution of a beam with fundamental (first peak) and harmonics (second peak) content.

\subsection{Energy resolution}

\begin{figure}[H]
    \centering
    \begin{subfigure}{0.32\textwidth}   
        \includegraphics[width=\textwidth]{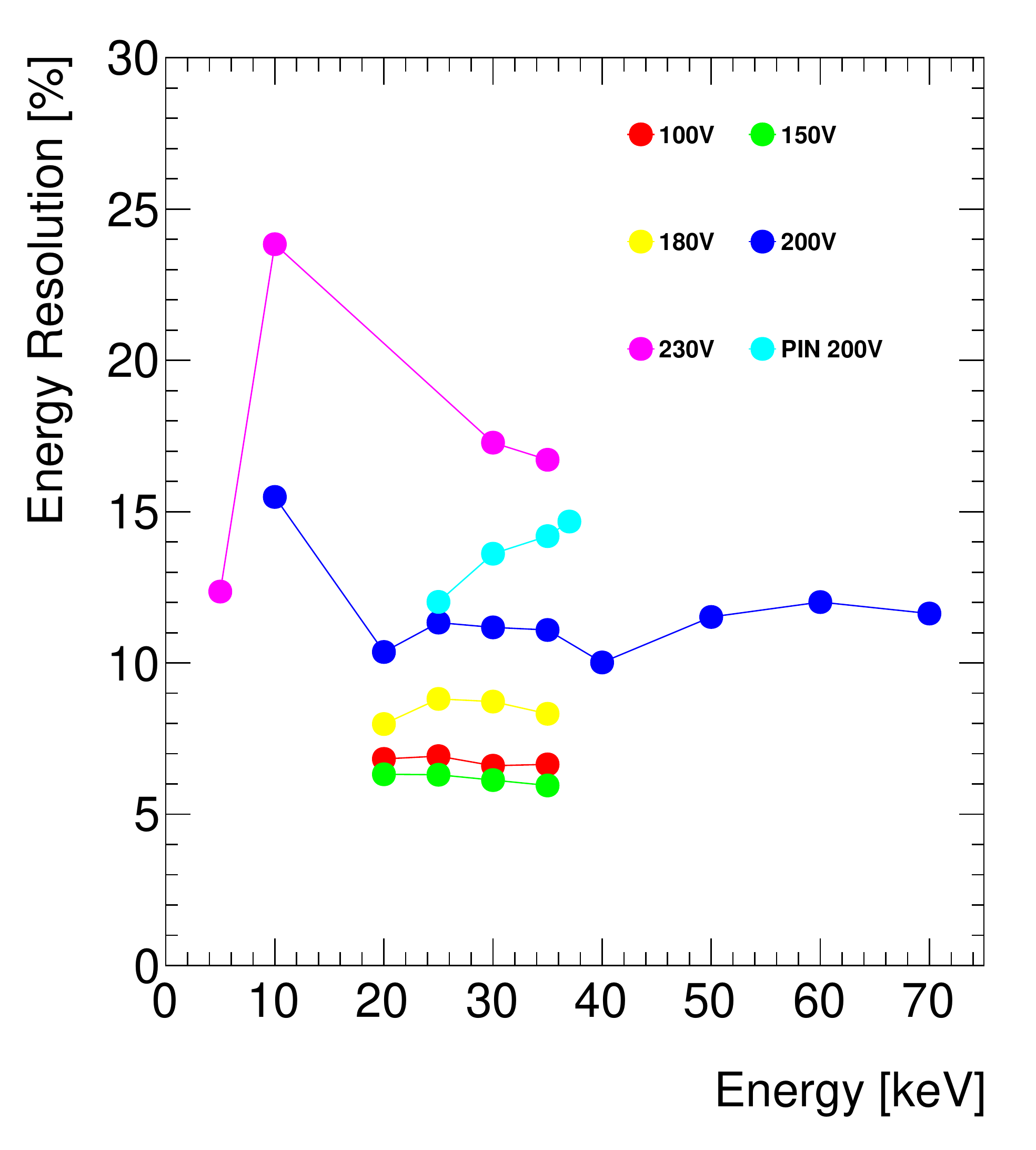}
        \caption{HPK PIN and type 3.1 LGAD}
        \label{fig:energyHPK31}
    \end{subfigure}
    \hfill  
    \begin{subfigure}{0.32\textwidth}   
        \includegraphics[width=\textwidth]{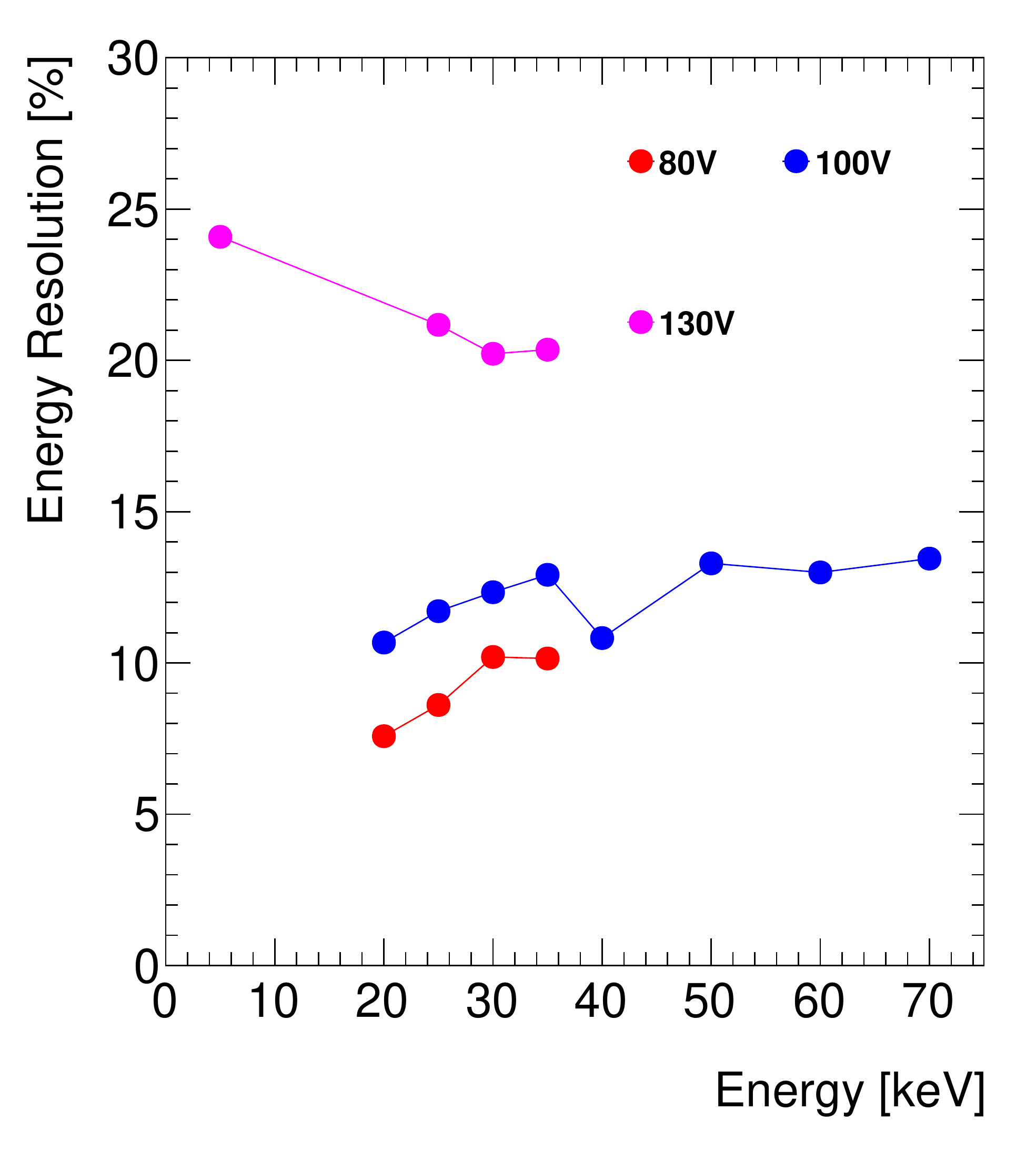}
        \caption{HPK type 3.2 LGAD}
        \label{fig:energyHPK32}
    \end{subfigure}
    \hfill  
    \vspace{0.5cm}
     \begin{subfigure}{0.32\textwidth}   
        \includegraphics[width=\textwidth]{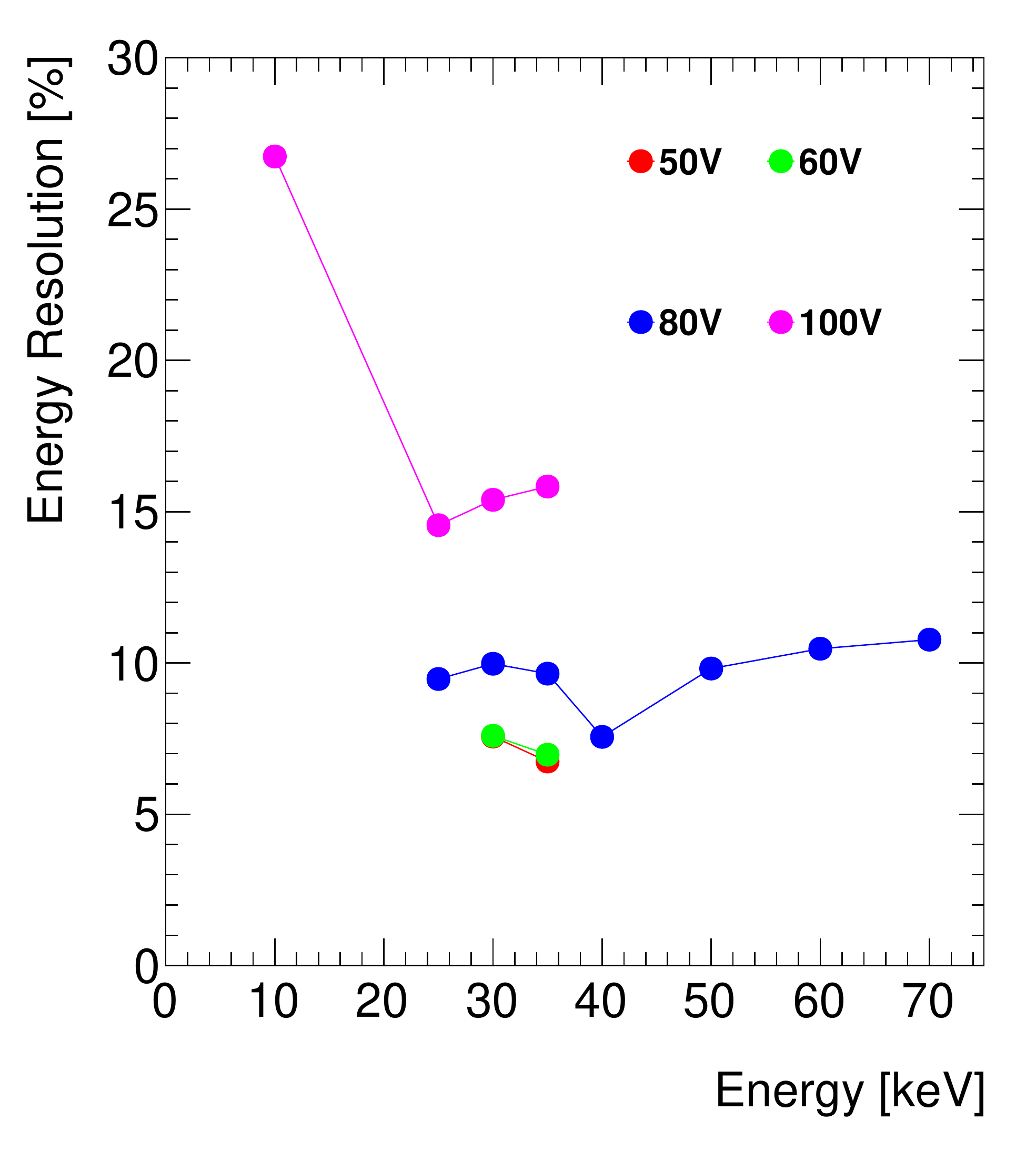}
        \caption{BNL 20um LGAD}
        \label{fig:energyBNL20}
    \end{subfigure}
        
    \caption{Energy resolution as $\frac{\sigma}{\mu}$ for the different bias voltages used on this test.}
    \label{fig:energy_resolution}
\end{figure}

The energy resolution of a sensor is calculated from the  $\pmax$ distribution of the sensor waveform pulses and defined as  $\frac{\sigma}{\mu}$, where $\sigma$ and $\mu$ are extracted from a Gaussian fit of the distribution (Fig.~\ref{fig:pmax_distribution}).
The estimated energy resolution is shown in Fig.~\ref{fig:energy_resolution} as a function of X-rays energy for several LGAD sensors at different bias voltages. Energies above that of \SI{35}{\kilo\electronvolt} are from harmonics. Fig.~\ref{fig:energy_resolution} shows that it is possible to achieve a resolution close to 7\% for energies above \SI{5}{\kilo\electronvolt} and there is a significant dependency of the energy resolution with the bias voltage (with higher voltages degrading significantly the energy resolution).


\subsection{Linearity of energy response}

\begin{figure}[H]
    \centering
    \begin{subfigure}{0.32\textwidth}   
        \includegraphics[width=\textwidth]{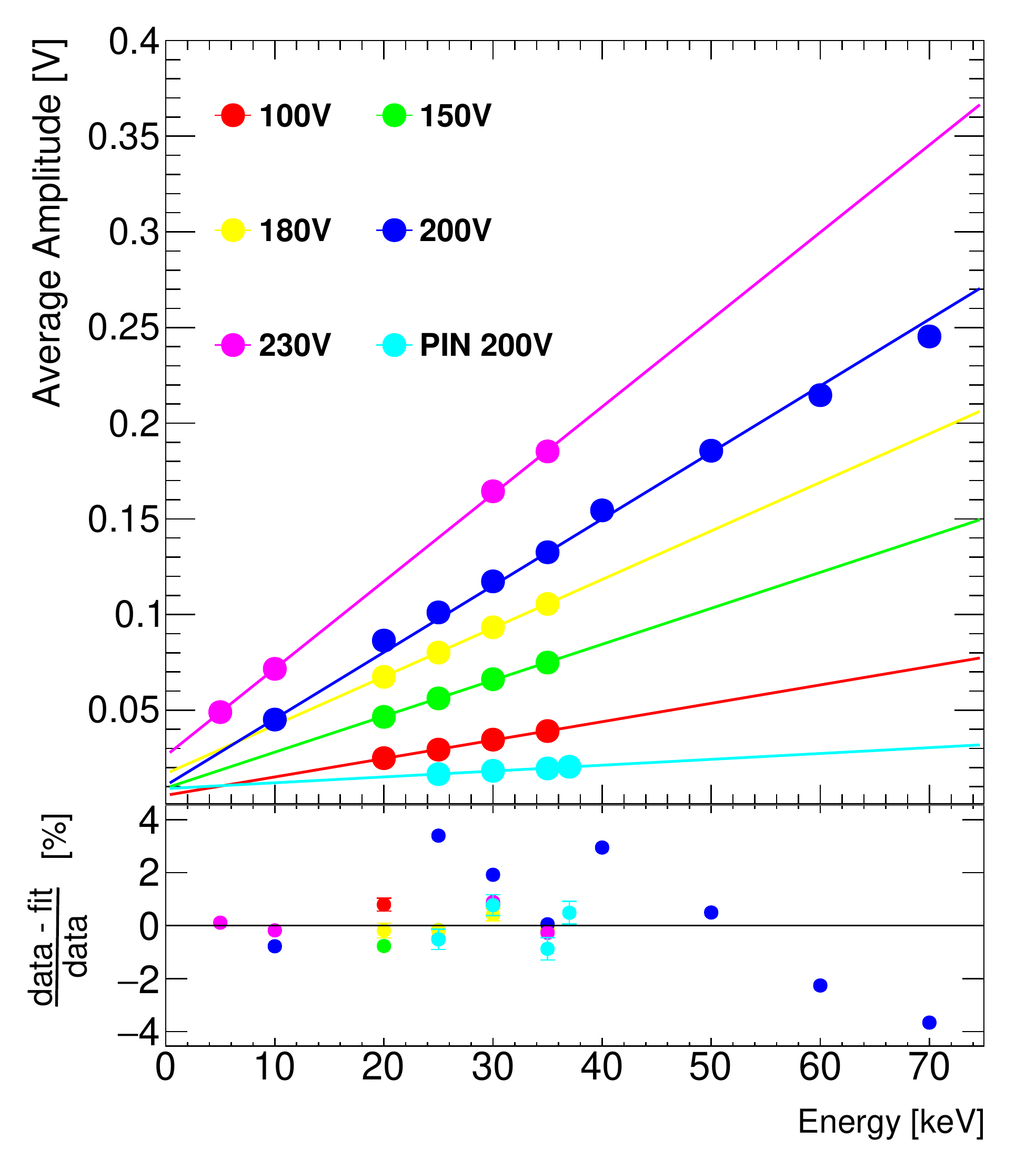}
        \caption{HPK PIN and type 3.1 LGAD}
        \label{fig:energy_linearity_HPK31}
    \end{subfigure}
    \hfill  
    \begin{subfigure}{0.32\textwidth}   
        \includegraphics[width=\textwidth]{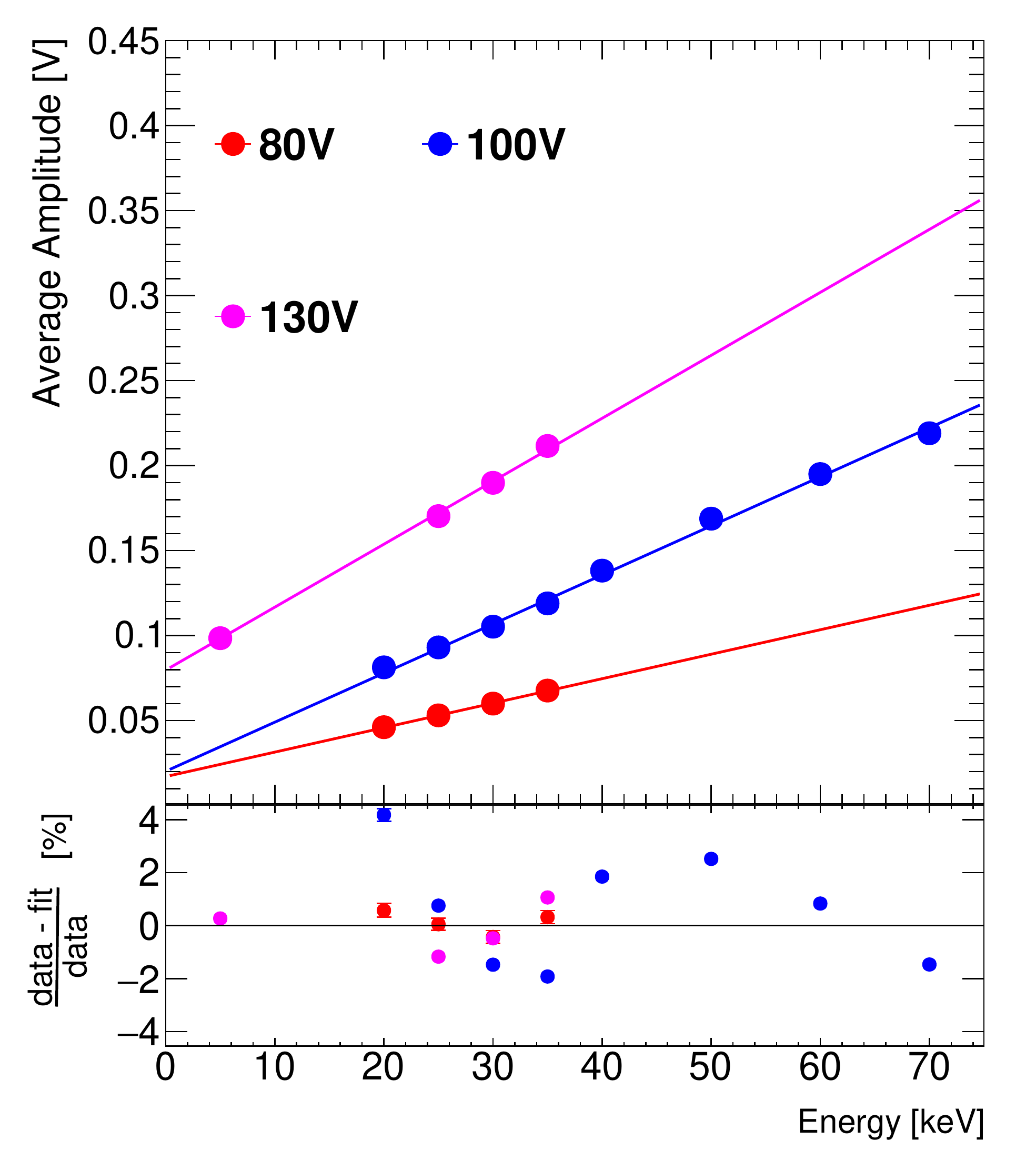}
        \caption{HPK type 3.2 LGAD}
        \label{fig:energy_linearity_HPK32}
    \end{subfigure}
    \hfill  
    \vspace{0.5cm}
    \begin{subfigure}{0.32\textwidth}   
        \includegraphics[width=\textwidth]{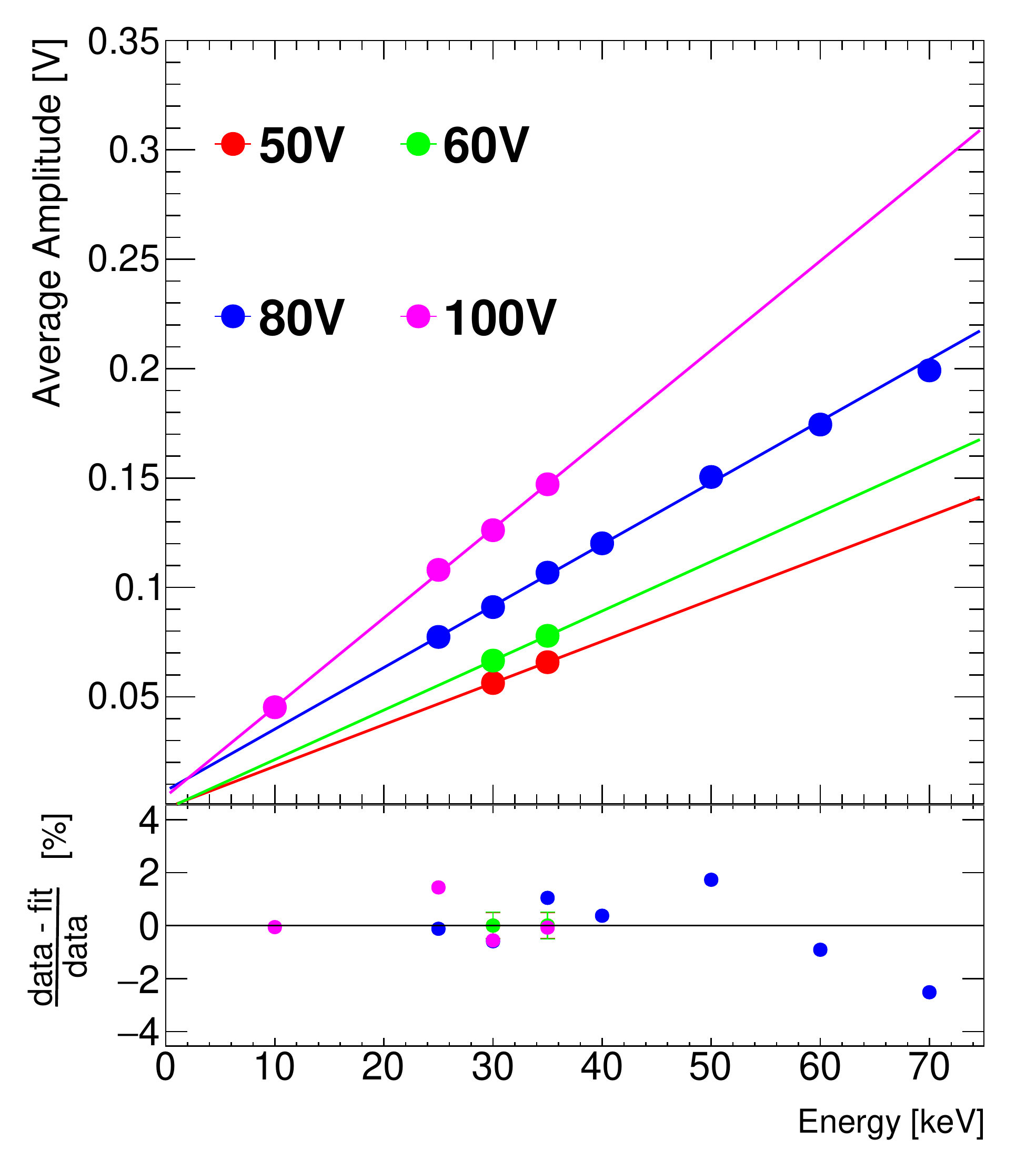}
        \caption{BNL 20um LGAD}
        \label{fig:energy_linearity_BNL20}
    \end{subfigure}
        
    \caption{Energy response of the 3 sensors tested. On the response plots, the points correspond to the average amplitude for each bias voltage and X-ray energy, the lines are linear fits on the data for the same bias voltage conditions, and the error bars are the Gaussian fit error on the average amplitude parameter on the $\pmax$ distribution. On the lower plots, the deviation around the linear fits shows a dispersion better than \SI{4}{\percent} for all the sensors tested. The error bars consider the average amplitude error and the energy response linear fit error.
    }
    \label{fig:energy_linearity}
\end{figure}

For many applications, it is highly desirable that the sensor responds linearly across a wide range of X-ray energies. Fig.~\ref{fig:energy_linearity} shows the relationship between energy and the sensor response taken as an average of $\pmax$  distribution and the plots of the residuals reveal a deviation of less than $4\%$ for the energy range from 5 to 70 keV. 
The gain of the LGAD sensors in Fig.~\ref{fig:energy_linearity} (obtained dividing the LGAD response with the PIN response) is different from the gain shown in Fig.~\ref{fig:LGADS_beta}.
This difference can be explained by the gain saturation of point-like charge deposition for an X-ray, as explained in Sec.~\ref{sec:TCAD_sat}.

\subsection{Time resolution}

Tests with LGADs using charged particles (MIP) have shown that these devices are able to achieve time resolution of tens of picoseconds~\cite{bib:LGAD} thanks to its internal charge multiplication mechanism. An evaluation of the timing response of LGADs to X-rays benefits from the very compact packet pulse of the SSRL (\SI{10}{\pico\second}) and from the synchronous nature of the SSRL beam.

\begin{figure}[H]
    \centering
    \includegraphics[width=0.6\linewidth]{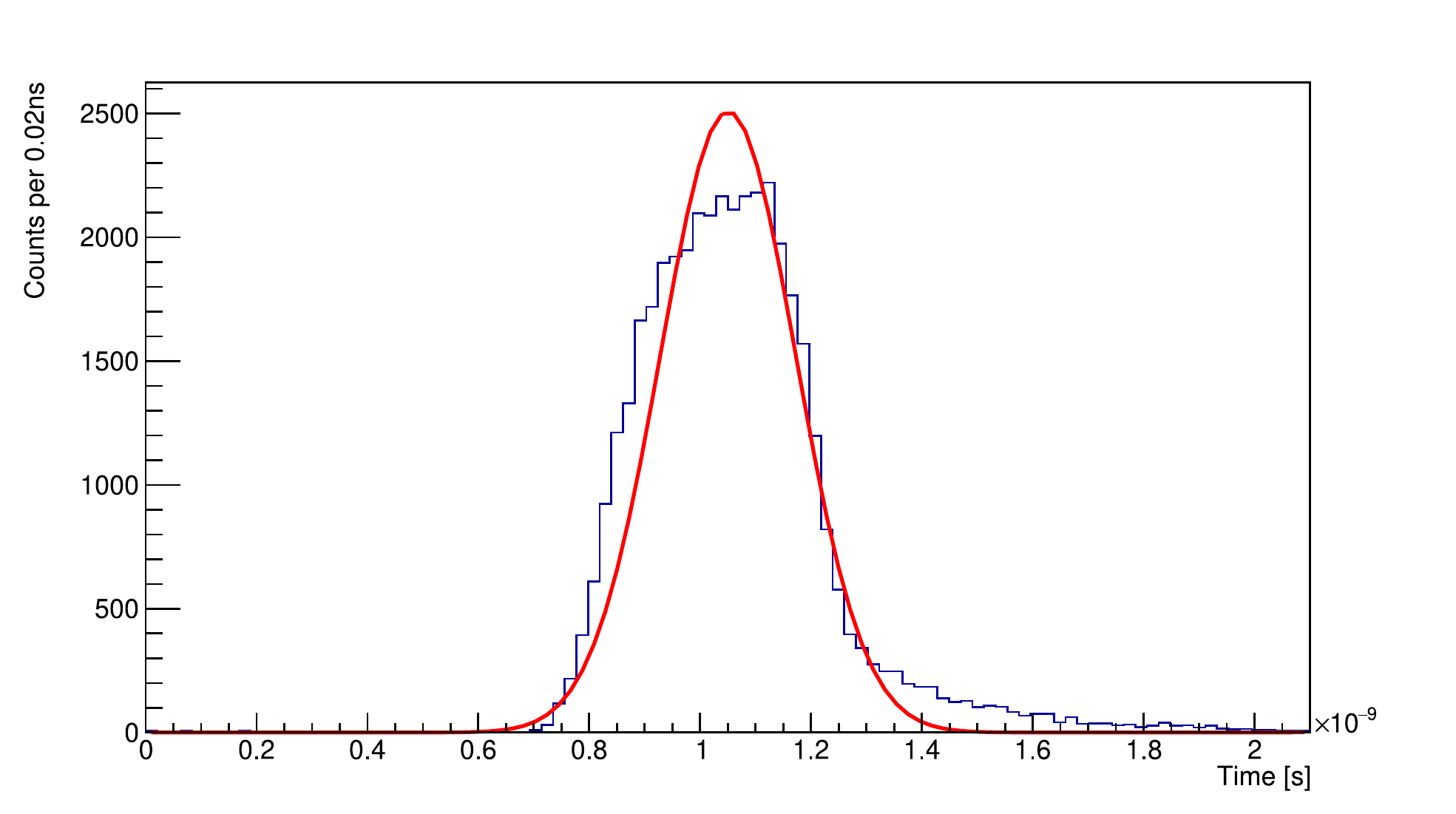}
    \caption{Time distribution of the waveform pulses calculated using CFD at 20\% and the timing signal of the SSRL beam. Data for HPK type 3.1 LGAD at \SI{200}{\volt} exposed to a beam of \SI{35}{\kilo\electronvolt} X-rays. The central value was artificially shifted to be centered at \SI{1.05}{\nano\second}.}
    \label{fig:CFD_distr}
\end{figure}

A Constant Fraction Discriminator (CFD) method with a 20\% fraction (the data presented in Fig.~\ref{fig:LGADS_beta} is for 50\% fraction, the time resolution in Sr$^{90}$ is marginally different between 20\% and 50\%) was then applied on the waveforms pulses using as a reference the precise $2.100 \pm 0.001$ ns\footnote{The resolution was estimated using the waveforms autocorrelation function.} inter-packet separation of SSRL to estimate the sensor time resolution for X-rays. The time resolution was calculated as the standard deviation of a Gaussian fit on the distribution of time as described (Fig.~\ref{fig:CFD_distr}).

\begin{figure}[H]
    \centering
    \begin{subfigure}{0.32\textwidth}   
        \includegraphics[width=\textwidth]{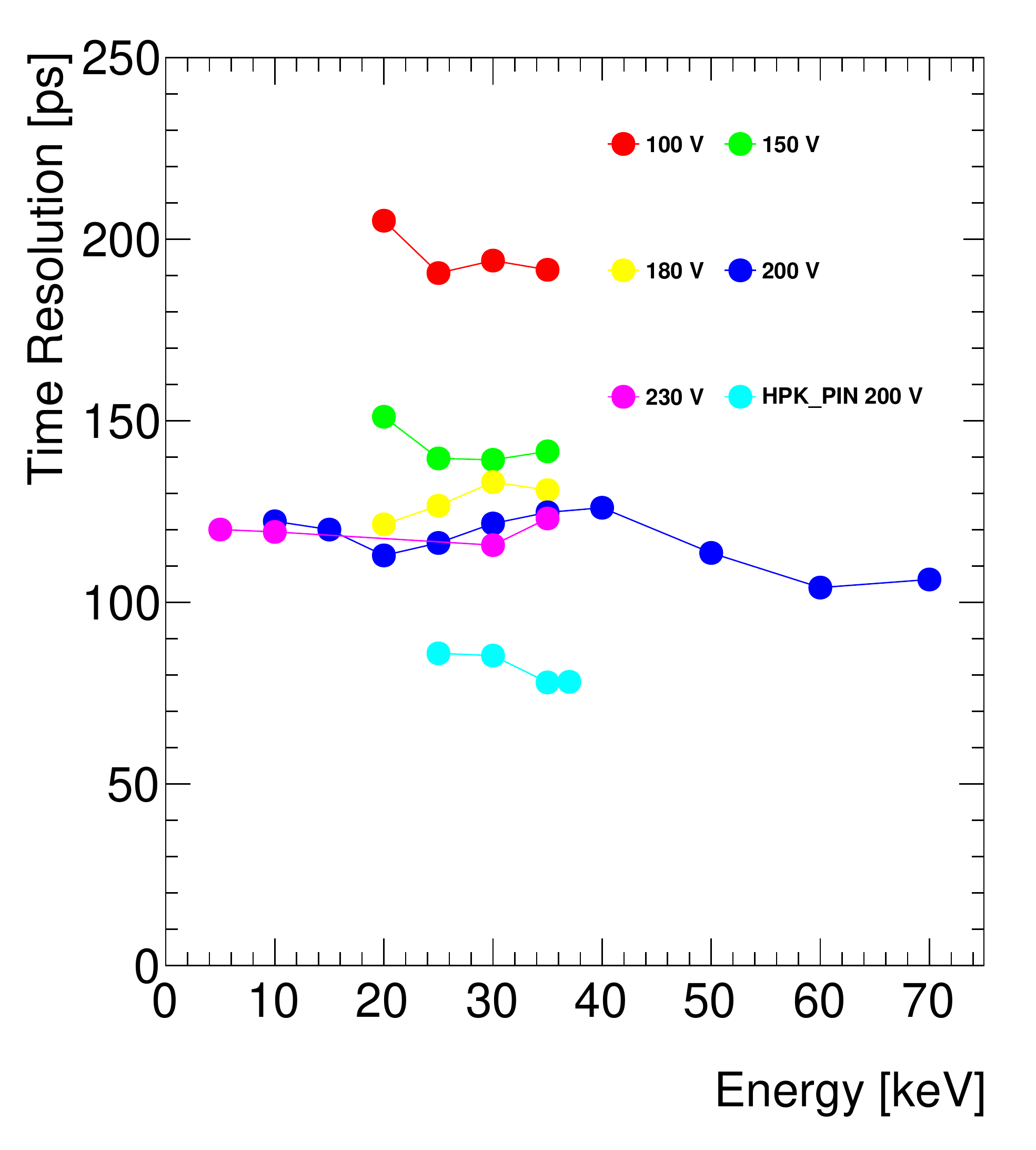}
        \caption{HPK PIN and type 3.1 LGAD}
        \label{fig:CFDHPK31}
    \end{subfigure}
    \hfill  
    \begin{subfigure}{0.32\textwidth}   
        \includegraphics[width=\textwidth]{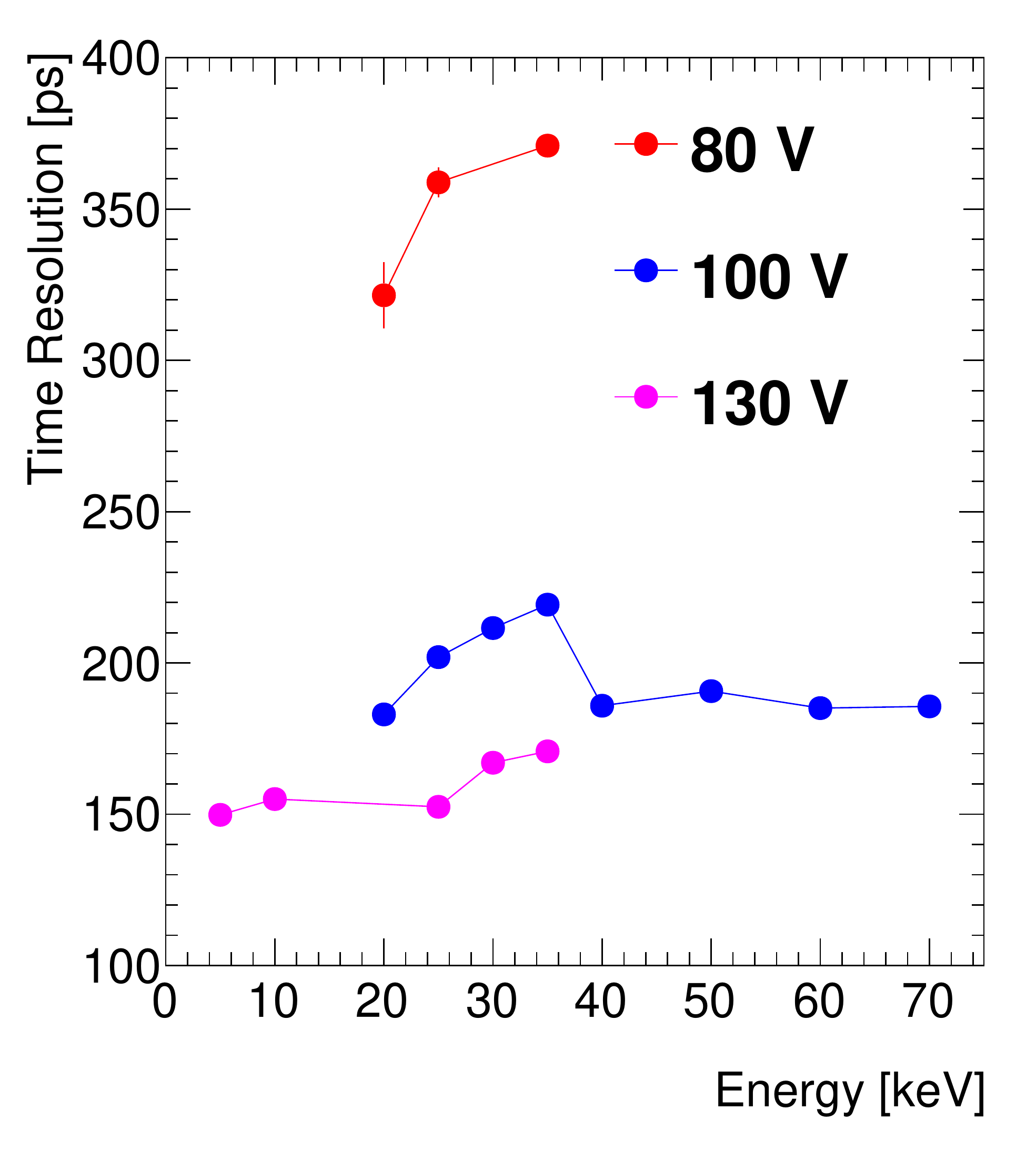}
        \caption{HPK type 3.2 LGAD}
        \label{fig:CFDHPK32}
    \end{subfigure}
    \hfill  
    \vspace{0.5cm}
     \begin{subfigure}{0.32\textwidth}   
        \includegraphics[width=\textwidth]{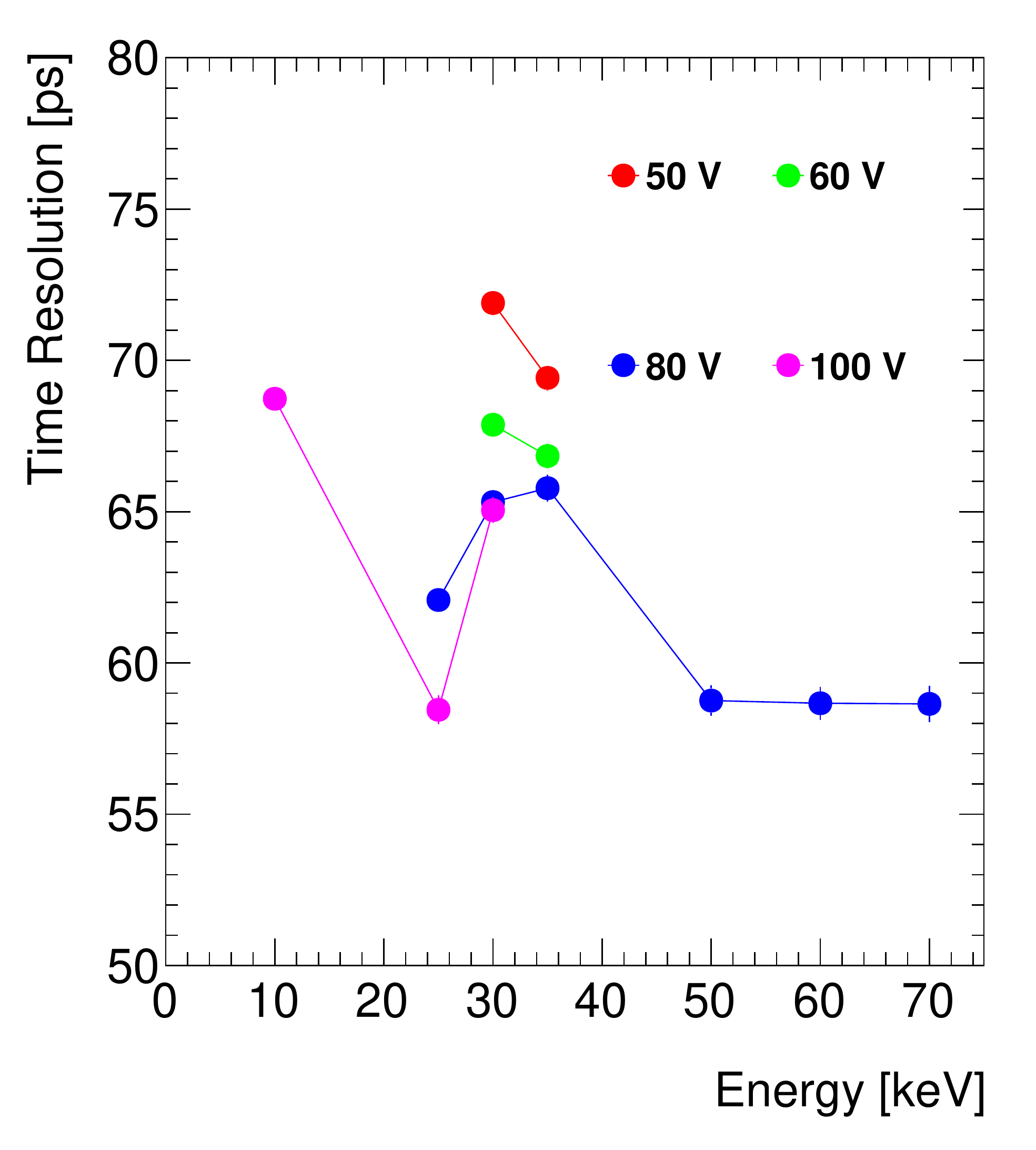}
        \caption{BNL 20um LGAD}
        \label{fig:CFDBNL20}
    \end{subfigure}
        
    \caption{Time resolution as a function of the X-rays energy for several LGAD sensor types and bias voltages. The timing was calculated from the waveform pulses using a Constant Fraction Discriminator with a 20\% fraction and the timing structure of the SSRL beam.}
    \label{fig:CFD_timing_resolution}
\end{figure}

%
%
Fig.~\ref{fig:CFD_timing_resolution} presents the time resolution as a function of the energy for several LGAD sensors at different bias voltages obtained using a CFD method (20\% fraction) for pulses from waveforms and the SSRL beam time structure information. A significant dependency with the bias voltage is observed, with higher voltages leading to a better time resolution. 
As seen in Fig.~\ref{fig:CFDHPK31}, the PIN device shows a better time resolution than the other \SI{50}{\micro\meter} LGAD devices. 
This is an effect of the X-ray's point-like current induced in the device. 
The current in the PIN device is instantaneous, while in the LGAD, the shape of the current induced by the X-ray interaction depends on the interaction depth, as shown in Fig.~\ref{fig:tcad:amp_depth_signal} for simulation and in Fig.~\ref{fig:pulses_normalized} in the actual data. The interaction depth causes a natural delay in the pulse time of arrival that cannot be corrected with a standard 20\% CFD algorithm normally used to reduce time-walk. 
Therefore, the time resolution of the PIN device is naturally better than the LGAD device's, even though the S/N is lower.

\subsection{Results}

The best achieved energy and time resolution for the LGAD and PIN geometries tested are shown in Tab.~\ref{tab:summary}. The trade-off between the best energy resolution for a lower bias voltage and the best time resolution for a higher bias can be seen for the 3 LGAD sensors.

\begin{table}[H]
\begin{tabular}{l|c|cc|cc|cc}
 
\multicolumn{1}{l}{} & HPK PIN & \multicolumn{2}{c}{HPK3.1} & \multicolumn{2}{c}{HPK3.2} & \multicolumn{2}{c}{BNL 20um}
\\ \hline
Bias V & \SI{200}{\volt} & \SI{150}{\volt} & \SI{230}{\volt} & \SI{80}{\volt} & \SI{130}{\volt} & \SI{50}{\volt} & \SI{100}{\volt} \\
Energy Resolution & \SI{14}{\percent} & \SI{6}{\percent} & \SI{17}{\percent} & \SI{10}{\percent} & \SI{20}{\percent} & \SI{6}{\percent} & \SI{16}{\percent} \\
Energy Response & \SI{19}{\milli\volt} & \SI{75}{\milli\volt} & \SI{185}{\milli\volt} & \SI{68}{\milli\volt} & \SI{211}{\milli\volt} & \SI{66}{\milli\volt} & \SI{147}{\milli\volt} \\
$\sigma_t$ CFD & \SI{78}{\pico\second} & \SI{141}{\pico\second} & \SI{123}{\pico\second} & \SI{371}{\pico\second} & \SI{171}{\pico\second} & \SI{69}{\pico\second} & \SI{65}{\pico\second}\\
\end{tabular}
\caption{Summary of energy and time resolution for the three tested sensors for the different bias voltages that yield the best energy and best time resolution for a \SI{35}{\kilo\electronvolt} X-ray beam energy.}
\label{tab:summary}
\end{table}

\section{Energy resolution for AC-LGADs}
\label{sec:ACLGAD}
\begin{figure}[H]
    \centering
    \begin{subfigure}{0.48\textwidth}  
    \includegraphics[width=\textwidth]
        {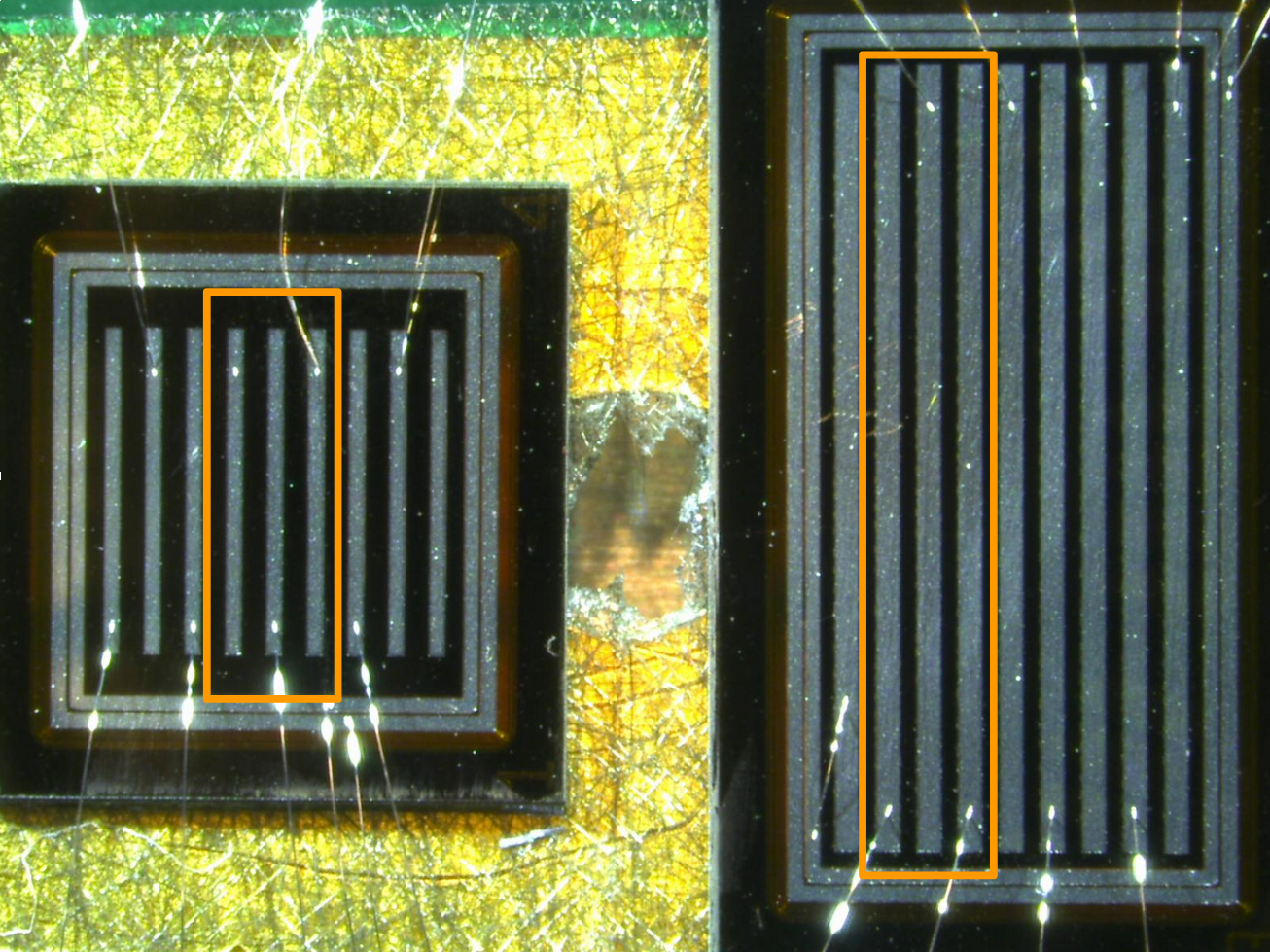}
        \caption{}
    \end{subfigure}
            \begin{subfigure}{0.48\textwidth}  
    \includegraphics[width=\textwidth]
        {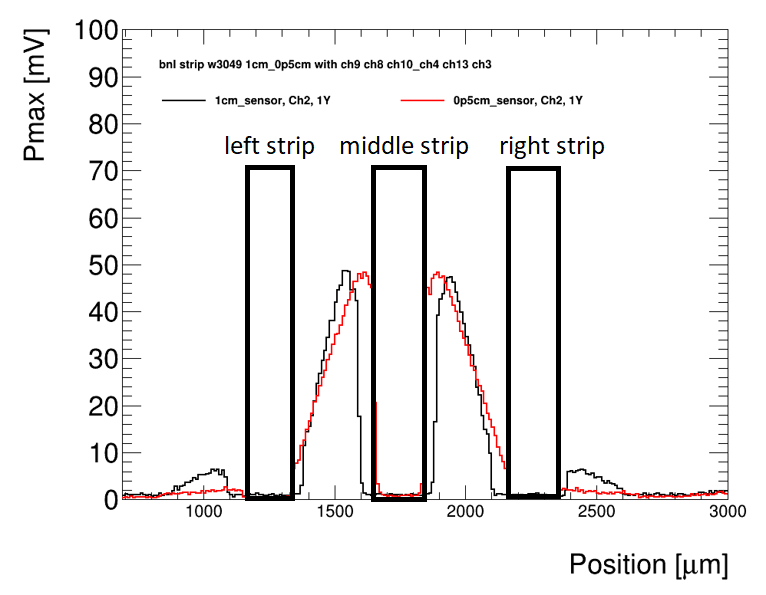}
        \caption{}
    \end{subfigure}
    \caption{(a) Picture of AC-LGAD with strip length $\SI{5}{\mm}$ and $\SI{10}{\mm}$. The strips used on the analysis are highlighted by the orange box. The sum of $\pmax$ from the three signal strips are used to measure the energy resolution. 
    (b) Response of the middle strip as a function of position perpendicular to the strip, AC-LGAD with strip length $\SI{5}{\mm}$ (red) and $\SI{10}{\mm}$ (black), data taken with an IR laser TCT setup.
    }
    \label{fig:ACLGAD_scans}
\end{figure}

The energy resolution was also measured for the two AC-LGAD devices in Fig.~\ref{fig:ACLGAD_scans}(a).
This measurement is of particular interest since for standard LGADs the granularity is limited to the millimeter scale, while for AC-LGADs a position resolution bellow \SI{50}{\micro\meter} is achievable.
The devices are AC-coupled, so the charge is shared between metal electrodes~\cite{Heller:2022aug} as shown in Fig.~\ref{fig:ACLGAD_scans}(b). 
Therefore, the energy response has to be measured by summing multiple strips.
The data in Fig.~\ref{fig:ACLGAD_scans}(b) was collected with an IR laser TCT station to emulate the effect of a minimum ionizing particle (MIP).
The surface of the device is scanned with a \SI{20}{\micro\meter} laser spot injecting the same energy of a MIP.
The response of the middle strip as a function of the direction perpendicular to the strip is shown in the plot. 
The zero response sections, highlighted by the black squares, correspond to the metal strips since the metal is not transparent to the laser light.
The charge is shared mainly between three strips in both devices, therefore the sum of the three highlighted strips in Fig.~\ref{fig:ACLGAD_scans}(a) will be used in the following.

\begin{figure}
    \centering
    \begin{subfigure}{0.48\textwidth}  
    \includegraphics[width=\textwidth]
        {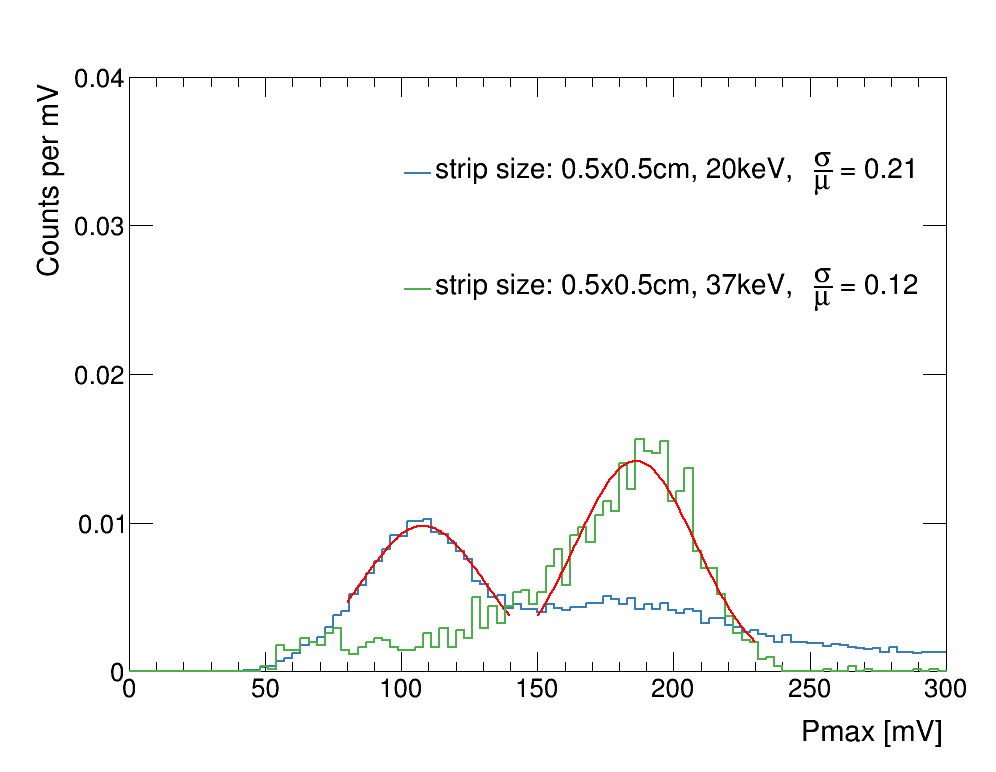}
        \caption{}
    \end{subfigure}
            \begin{subfigure}{0.48\textwidth}  
    \includegraphics[width=\textwidth]
        {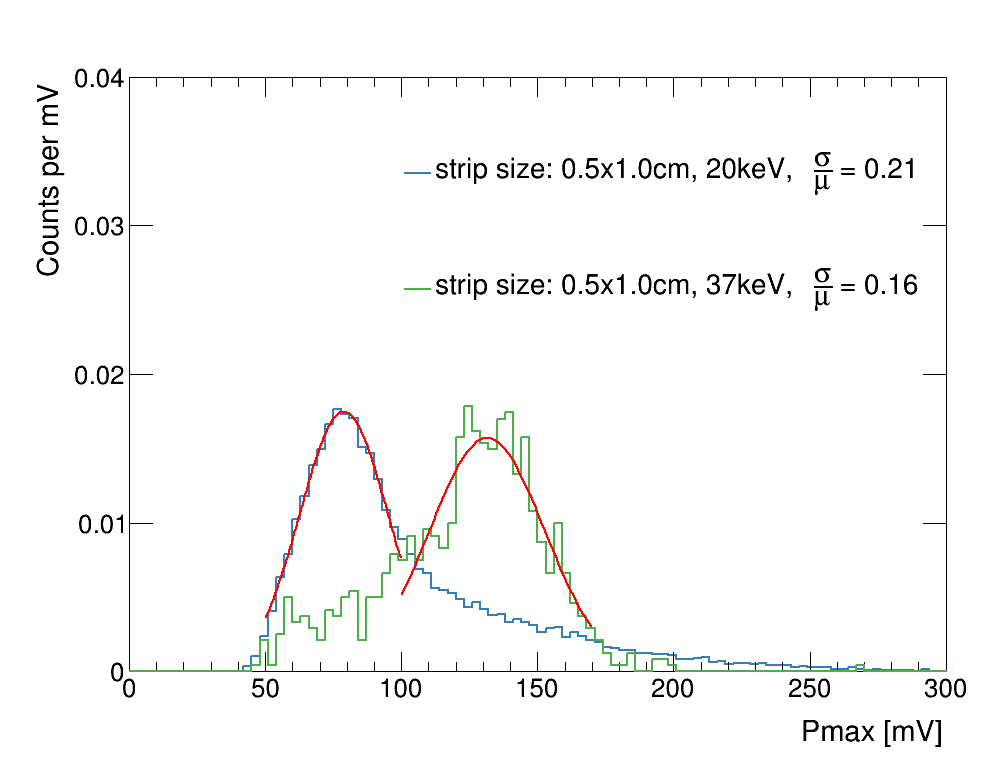}
        \caption{}
    \end{subfigure}
    \caption{
        $\pmax$ distributions for the sum of three selected strips for 5~mm long (a) and 10~mm long (b) strip sensors with superimposed energy resolution fit. A selection of 40~mV is applied to the middle strip in both cases to remove events not centered in the three strips. The beam was illuminating more than the 5~mm sensor; this explains the higher tail due to the increased number of double photon events.
    }
    \label{fig:ACLGAD_energy}
\end{figure}

Since the X-ray beam was 25~mm~x~1~mm in size, no position information was available.
A simple analysis was applied to measure the energy resolution, although not with the highest precision due to the lack of position information. 
A group of three adjacent strips is selected as shown in Fig.~\ref{fig:ACLGAD_scans}(a), and the middle strip is required to have the highest signal of the entire sensor.
The two neighboring strips need to have a signal lower than the middle strip, and their $\tmax$ is required to be within \SI{600}{\pico\second}.
The $\pmax$ of the three strips is then summed to measure the deposited energy as seen in Fig.~\ref{fig:ACLGAD_energy}.
The $\pmax$ is used instead of the pulse area because the waveform should have zero area in AC-LGADs since it’s an AC-coupled signal.
Then, akin to the studies performed in the previous sections, the $\pmax$  distribution is fitted with a Gaussian to measure central value and resolution.
This study was only made with two energies and at a low rate (baseline only) to reduce the number of multiple hits on the large sensor.
The energy response of the two sensors is proportional to the X-ray energy with a resolution between 12\% and 21\%.
The \SI{5}{\milli\meter} device shows a slightly better energy resolution at \SI{37}{\keV}, probably related to the reduced charge sharing between strips.

\FloatBarrier

\section{Conclusions}
\label{sec:conclusions}

X-ray detection of silicon devices (LGADs and PIN) was studied at the Stanford Light Source (SSRL). 
The tested thin devices easily resolved in time the 500~MHz repetition rate of the beam line.
The energy resolution of the tested LGADs was measured to be between 6\% and 20\%, depending on the X-ray energy and the gain of the device. 
The best energy resolution (6\%) was obtained with LGADs operated at a voltage with a gain of roughly ten.
At high voltages, it was found that the gain fluctuations negatively impact the energy resolution.
The PIN (no gain) device energy resolution is 2-3 worse than the best LGAD energy resolution.
The gain of the LGADs is lower for X-rays than for MIP because of gain saturation effects. 
The effect was seen in TCAD simulations and confirmed in the data.
A preliminary estimation of the energy response of AC-LGADs was made, and it showed an energy resolution of 12-21\%, somewhat higher than the DC-LGAD counterpart; the degradation is expected given the AC nature of the device and the charge-sharing mechanism. However, the broad beam didn't allow for precision studies as a function of position. 

The time resolution was estimated to be between 50~ps and 200~ps after time-walk correction; this value is significantly worse than the case of minimum ionizing particles (usually between 20~ps to 50~ps). 
This is explained by the different charge deposit profiles between minimum ionizing particles and X-rays (point-like); the effect was simulated using TCAD Sentaurus.
The time delay due to charge deposition depth was also observed in data using binned averaged waveforms, although with a higher gain variation than the one predicted in the simulation.
For each device, the best time resolution is achieved at the maximum voltage.
As expected, a better time resolution is achieved using the \SI{20}{\micro\meter} device, where the total drift time is shorter.
The PIN showed better time resolution than same-thickness LGAD devices because of the lack of dependency on the X-ray's interaction depth caused by the gain mechanism.

{
\small
\acknowledgments
This work was supported by the United States Department of Energy, grant DE-FG02-04ER41286.
CACTUS DJ-LGAD SBIR.
Use of the Stanford Synchrotron Radiation Lightsource, SLAC National Accelerator Laboratory, is supported by the U.S. Department of Energy, Office of Science, Office of Basic Energy Sciences under Contract No. DE-AC02-76SF00515.
The group from USP acknowledges support from FAPESP (grant 2020/04867-2) and CAPES.
\bibliography{bib/TechnicalProposal,bib/hpk_fbk_paper,bib/HGTD_TDR,bib/data_analysis, bib/others}
}

\end{document}